\documentclass[a4paper,fleqn,numbers]{cas-dc}
\usepackage{etoolbox} 
\usepackage{xspace}   
\usepackage[numbers]{natbib}
\usepackage{graphicx} 

\usepackage{capt-of}   
\usepackage{float}
\usepackage{rotating}
\usepackage{pifont}
\usepackage{booktabs}
\usepackage{array}
\usepackage{xcolor}
\usepackage{placeins}
\usepackage{graphicx}
\usepackage{amsmath}
\usepackage{booktabs}
\usepackage{placeins}
\usepackage{subcaption}
\usepackage{comment}
\usepackage{soul}
\newcommand{\cmark}{\ding{51}}
\newcommand{\xmark}{\ding{55}}

\def\tsc#1{\csdef{#1}{\textsc{\lowercase{#1}}\xspace}}
\tsc{WGM}
\tsc{QE}
\tsc{EP}
\tsc{PMS}
\tsc{BEC}
\tsc{DE}

\begin{document}

\shorttitle{Autoscaling in Serverless Computing}

\shortauthors{Author et~al.}

\title[mode=title]{A Unified Queueing-Based Framework for Comparative Evaluation of Serverless Autoscaling Strategies under Bursty Workloads}

\author[1]{Zahra Zafarzade}
\ead{z.zafarzade11@umail.umz.ac.ir}

\author[1]{Ehsan Ataie}
\cormark[1]
\ead{ataie@umz.ac.ir}

\affiliation[1]{organization={Department of Computer Engineering, University of Mazandaran},
	city={Babolsar},
	country={Iran}}

\cortext[cor2]{Corresponding author}


\begin{abstract}
	Serverless computing platforms must preserve quality of service (QoS) while adapting to highly bursty and unpredictable workloads. To cope with workload fluctuations while satisfying  QoS requirements and service-level agreements (SLAs), these platforms rely on automatic resource scaling. Commercial function-as-a-service (FaaS) platforms employ different autoscaling strategies to dynamically adjust the number of active function instances and the available processing capacity. In this work, three representative autoscaling strategies are modeled and compared: scale-per-request (SPR), concurrency-level scaling (CLS), and metric-based scaling (MBS).	
	To enable systematic evaluation independent of proprietary platform-specific implementations, we develop a unified modeling framework that combines analytical queueing network (QN) modeling with queueing Petri nets (QPN) implemented in Java Modeling Tools (JMT) to evaluate autoscaling behavior under two-phase hyper-exponential ($H_2$) arrivals. Evaluation is conducted using four system-level metrics: throughput, mean response time, system population, and system power. The framework captures dynamic function instance (FI) activation, concurrency constraints, and queue buildup under saturation. Comparative analysis reveals a clear trade-off between peak performance and stability: SPR provides the lowest delay under moderate load but becomes fragile near capacity limits; CLS is governed by concurrency tokens and is highly sensitive to configuration under bursty conditions; whereas MBS provides the most stable QoS across the workload range by keeping utilization below a critical region and preventing explosive queue growth under critical load.
\end{abstract}


\maketitle

\section{Introduction}

Serverless computing has emerged as a major paradigm in cloud systems. Under the
function-as-a-service (FaaS) model, services are executed in an event-driven manner and deployed
without direct management of the underlying infrastructure
\cite{hong2024analysis, tari2024auto}. To execute these functions, the platform creates and manages isolated execution environments, referred to in this study as function instances (FIs). An FI provides the runtime environment and the computational resources required to process function invocations \cite{agarwal2024dynamic, fan2024puffer}. Depending on the platform architecture, these execution environments may be implemented using containers, virtual machines, or lightweight microVM-based isolation mechanisms \cite{agarwal2024dynamic, li2024secure, karamzadeh2024lightweight}. These implementation details are typically hidden from
the user, while the FI acts as the practical unit through which processing capacity is
provided and adjusted.

Resource provisioning is therefore handled by the platform, which reacts to workload variation
by activating, reusing, or retiring FIs. Depending on the platform and its concurrency
configuration, an FI may process a single request or multiple concurrent requests. While this
abstraction reduces operational burden, it also makes quality of service (QoS), especially
under dynamic demand, strongly dependent on the design and tuning of autoscaling mechanisms
\citep{hong2024analysis, tari2024auto, Alharthi2024Auto-Scaling}.

In real deployments, request streams are often bursty and difficult to anticipate, and such variability can amplify queueing effects as systems approach saturation~\citep{Alharthi2024Auto-Scaling, Meng2025CATScaler, Jeong2025ARAScaler}.
FaaS platforms must therefore increase capacity within short time windows while facing practical
constraints such as instance activation delays, per-instance concurrency limits, and finite resources~\citep{Ebrahimi2024Cold, Cai2025SLO-Aware}.
Under such conditions, delayed or unstable scaling decisions may prevent FaaS platforms from responding effectively to workload fluctuations. This limitation can drive the system toward sustained high utilization, where cold-start overheads, resource contention, and queue accumulation become more pronounced, ultimately resulting in degraded user-perceived latency~\citep{benedetti2025qlearning, Cai2024Incendio:,Jeong2025ARAScaler}.

Existing studies have investigated autoscaling and performance modeling for serverless and
cloud-native platforms using analytical methods, simulation, and measurement-based evaluation~\citep{hong2024analysis, Quattrocchi2024Autoscaling}.
However, comparing different autoscaling strategies under identical conditions remains challenging because workload characteristics and
platform constraints can strongly affect their observed behavior.
In addition, commercial FaaS platforms provide
limited visibility into their internal scaling mechanisms, making
detailed analysis and independent validation more difficult
\citep{Kulkarni2025Characterizing}.

To address this gap, a unified modeling framework is presented for studying autoscaling
behavior in FaaS platforms. Queueing network (QN) theory is combined with
queueing petri nets (QPN) implemented in the java modeling tools (JMT),
enabling a consistent representation of capacity adaptation and system
behavior under increasing load. Bursty demand is represented using a two-phase hyper-exponential ($H_2$)
arrival process, allowing controlled evaluation under high-variability traffic. Within this framework,
three autoscaling families are modeled and compared: scale-per-request (SPR), concurrency-level scaling
(CLS), and metric-based scaling (MBS).

The main contributions of this work are summarized as follows:

\begin{enumerate}
	
	\item A unified modeling and evaluation framework is developed for FaaS autoscaling strategies by combining QN models with QPN-based representations. This framework provides a transparent and reproducible foundation for analyzing autoscaling behavior.
	
	\item The proposed QPN models explicitly represent operational constraints of FaaS platforms, including FI availability, per-instance concurrency limits, and queue accumulation under constrained processing capacity.
	
	\item A burst-aware workload representation based on an $H_2$ arrival process is incorporated to investigate autoscaling behavior under variable traffic conditions.
	
	\item To the best of our knowledge, this work presents the first comprehensive analytical comparison of three FaaS autoscaling paradigms, including SPR, CLS, and MBS scaling, within a unified framework. Furthermore, the MBS strategy combines utilization monitoring phases with adaptive routing decisions in a threshold-driven metric-based scaling approach.
	
	\item A QoS-oriented performance assessment approach is established to analyze the impact of autoscaling decisions on service quality. By considering throughput, mean response time, in-system population, and system power, the approach reveals the trade-offs between scaling responsiveness and system stability.
	
\end{enumerate}

The remainder of this paper is organized as follows.
Section~\ref{related_work} reviews related work.
Section~\ref{sec:background_system} introduces the analytical
modeling formalisms used in this study.
Section~\ref{sec:desc_system} describes the system under study,
including the workload characteristics and modeling assumptions.
Section~\ref{sec:Proposed} presents the proposed autoscaling models
and evaluation metrics.
Section~\ref{sec:performance_results} reports and discusses the
performance evaluation results.
Finally, Section~\ref{sec:conclusion} concludes the paper and
outlines directions for future work.

%
\section{Related Work}
\label{related_work}
Many studies have investigated resource management, capacity planning, and performance optimization in microservices and serverless computing environments. These studies cover a wide range of autoscaling mechanisms and evaluation approaches, including empirical, control-theoretic, analytical, and stochastic modeling methods.

To provide a structured overview, the related work is divided into two main groups. Section~\ref{related1} reviews autoscaling mechanisms in microservices and serverless platforms, while Section~\ref{related2} examines analytical and stochastic modeling approaches to autoscaling behavior and system performance.

\subsection{Autoscaling in Serverless and Microservices}
\label{related1}

Tran and Kim \cite{tran2024concurrent} proposed a proactive hybrid autoscaling method for quota-constrained Knative systems, combining traffic prediction and deep reinforcement learning for concurrent services sharing a resource quota. Their experiments report fewer latency SLO violations and improved quota compliance compared with conventional Knative autoscaling, highlighting the need to coordinate scaling decisions across services.

Tran and Kim \cite{tran2024optimized} developed a hybrid autoscaling approach for Knative-based serverless edge computing. It combines horizontal and vertical scaling by adjusting instance counts, CPU allocation, and request concurrency using predicted traffic and service profiles. The implementation extends Kubernetes operators without modifying Knative and reduces resource usage while maintaining latency SLOs, although it depends on accurate workload prediction and profiling.

Shafi et al. \cite{Shafi2025CEMA} proposed CEMA, a predictive multi-layer autoscaling framework for microservices. It dynamically adjusts container and VM counts according to workload demand and migrates containers from underutilized VMs to reduce resource waste and infrastructure cost. Evaluated using real-world workload traces, CEMA focuses on containerized resource management rather than analytical comparison of FaaS scaling strategies.

Zafeiropoulos et al. \cite{Zafeiropoulos2022RLAutoscaling} developed reinforcement learning (RL)-assisted autoscaling mechanisms based on Q-learning, DynaQ+, and deep Q-learning. The agents make scaling decisions using resource and QoS metrics and are evaluated in both Kubeless deployments and simulated environments. The results support RL-based adaptation under dynamic workloads, although more complex agents require substantial training.

Khaleq and Ra \cite{khaleq2023intelligent} extended Kubernetes HPA with an RL-based autoscaling module for microservices. Their method selects CPU or memory metrics according to workload demand to improve response time. Experiments on Google Kubernetes Engine (GKE) with a Twitter analytics application show lower latency than default HPA, with the evaluation primarily addressing horizontal scaling.

Benedetti et al. \cite{benedetti2025qlearning} proposed a Q-learning-based autoscaling method for serverless functions in edge computing. It dynamically adjusts HPA configurations using Azure invocation patterns to improve latency-based SLA compliance while limiting unnecessary resource allocation. Experimental results indicate lower resource consumption and fewer request losses than baseline HPA configurations.

Mampage et al. \cite{mampage2025deep} introduced an A3C-based multi-agent RL approach that jointly optimizes horizontal and vertical scaling in multi-tenant serverless applications. The method balances application response time and infrastructure cost through configurable optimization priorities. Evaluations using Azure traces and a Kubeless-based Kubernetes testbed show that different priorities favor either lower latency or reduced cost.

Li et al. \cite{li2025energy} presented EAES, an energy-aware elastic scaling algorithm for Kubernetes microservices that coordinates feedforward and feedback control. It improves container provisioning by releasing excess containers and accounting for energy consumed by idle resources. Experiments report a 15.34\% reduction in cluster energy consumption compared with baseline scaling methods while maintaining SLA requirements.

Abdullah et al. \cite{abdullah2025dynamic} proposed an RL-based autoscaling approach that dynamically selects heterogeneous VM types to balance application performance and infrastructure cost. Evaluated using five real-world workload traces, including ClarkNet, NASA, Wikipedia, Calgary, and FIFA World Cup, it processes 10.49\% more requests on average and achieves 7.89$\times$ fewer SLO violations than the evaluated baselines.

\subsection{Autoscaling with Analytical and Stochastic Modeling}
\label{related2}

Ship et al. \cite{ship2024optimizing} developed an optimization-based serverless autoscaling method using a fluid approximation and Separated Continuous Linear Programming (SCLP). Their approach jointly determines per-replica resource allocation and the required number of replicas over time to reduce response times and failed requests. Python/SimPy-based simulations demonstrate improvements over threshold-based autoscaling under changing workloads.

Zafarzade and Ataie \cite{zafarzade2025capacity} proposed an analytical capacity planning approach for a microservices-based image classification application using Extended Queueing Networks. By analyzing utilization and performance at the function-station and system levels, their model identifies bottlenecks and guides changes in the number of FIs to avoid overutilization under varying workloads.

Wu et al. \cite{wu2025cloudnativesim} introduced CloudNativeSim, a CloudSim-based toolkit for modeling and simulating cloud-native microservice applications. The framework represents service dependencies and supports configurable scheduling and scaling policies. Validation against a reference deployment reports over 94.5\% accuracy in response-time simulation, demonstrating its suitability for evaluating alternative scaling policies.

F'{e} et al. \cite{fe2022performance} proposed a stochastic Petri net (SPN) model for performance--cost analysis of cloud autoscaling. It captures VM instantiation and termination and evaluates throughput, response time, and provisioning cost across configurations. Combined with GRASP optimization, the model identifies cost-effective settings subject to SLA constraints and is validated against 18,000 real-testbed samples.

Gias et al. \cite{gias2019atom} presented ATOM, a model-driven autoscaling framework for microservices based on runtime layered queueing network (LQN) models. It jointly adjusts replica counts and CPU shares, combining horizontal and vertical scaling. Under heavy workloads, the study reports 30--37\% higher throughput than static-rule controllers, although its target environment is microservices rather than FaaS.

Luan and Shen \cite{luan2024minimize} proposed an SLO-aware optimization method for containerized microservices combining layered queueing networks (LQNs), deep neural networks, and particle swarm optimization. Their approach seeks to reduce CPU resource costs while satisfying service-level objectives, with an emphasis on resource allocation in containerized applications.

Gao et al. \cite{gao2024performance} developed analytical performance models for distributed data processing in microservices. The models account for data-access workflows, caching behavior, and queueing contention to evaluate performance and guide service/data placement. Their primary focus is performance prediction and placement optimization rather than runtime autoscaling decisions.

Lim et al. \cite{lim2026sinas} introduced S-iNAS, a stochastic reward net (SRN)-based framework for analyzing time- and event-based scaling in distributed storage systems. Their results indicate that event-based scaling can reduce request latency compared with time-based policies, although it may trigger more frequent resource reconfiguration. The study addresses storage-layer rather than FaaS autoscaling.

Ataie et al. \cite{ataie2022modeling} proposed a stochastic activity network (SAN)-based model for evaluating dispatching policies in IaaS cloud data centers. The model analyzes performance, availability, and power consumption and is validated against CloudSim. Comparisons across dispatching policies illustrate its analytical capabilities, while the primary focus remains resource dispatching rather than autoscaling control.

Merkouche et al. \cite{merkouche2023petri} proposed a Petri net-based formal model for microservice autoscaling. The study explores the formal representation of microservice scaling behavior rather than a performance comparison of multiple serverless autoscaling strategies.

Jawaddi et al. \cite{jawaddi2025energy} investigated energy-aware Kubernetes microservice autoscaling through four Markov Decision Process (MDP) variants and probabilistic model checking. Their models incorporate scaling constraints inspired by Kubernetes HPA and support design-time evaluation of host energy consumption and service-level requirements under different scaling policies.

Although previous studies have advanced autoscaling mechanisms and performance modeling, these topics are often investigated separately. Existing work commonly evaluates a particular scaling policy or applies analytical and simulation-based models to a specific system. Relatively few studies compare distinct FaaS autoscaling strategies under common workload and system assumptions. Moreover, bursty arrivals and practical constraints, including per-instance concurrency, capacity management, and queue buildup near saturation, are not always considered together. A unified framework can therefore support more consistent and reproducible comparisons under dynamic workloads.

\begin{table*}[htbp]
	\centering
	\caption{Comparison of Existing Studies and the Proposed Framework}
	\label{tab:related_work_comparison}
	
	\renewcommand{\arraystretch}{1.1}
	
	\resizebox{\textwidth}{!}{
		\begin{tabular}{lccccccccc}
			\toprule
			
			\textbf{Study} &
			\makecell{\textbf{Analytical/}\\\textbf{Formal}\\\textbf{Modeling}} &
			\makecell{\textbf{Simulation-}\\\textbf{Based}\\\textbf{Evaluation}} &
			\makecell{\textbf{Cross-Strategy}\\\textbf{Comparative}\\\textbf{Modeling}} &
			\makecell{\textbf{Serverless}\\\textbf{Environment}} &
			\makecell{\textbf{Bursty}\\\textbf{Workload}} &
			\makecell{\textbf{Explicit Request}\\\textbf{Routing/}\\\textbf{Dispatching}} &
			\makecell{\textbf{Request-}\\\textbf{Concurrency}\\\textbf{Constraint}} &
			\makecell{\textbf{Horizontal}\\\textbf{Scaling}} &
			\makecell{\textbf{Vertical}\\\textbf{Scaling}}
			\\
			
			\midrule
			
			Tran and Kim \cite{tran2024concurrent} &
			\xmark & \xmark & \xmark & \cmark & \cmark & \xmark & \xmark & \cmark & \cmark \\
			
			Tran and Kim \cite{tran2024optimized} &
			\xmark & \xmark & \xmark & \cmark & \cmark & \xmark & \cmark & \cmark & \cmark \\
			
			Shafi et al. \cite{Shafi2025CEMA} &
			\xmark & \cmark & \cmark & \xmark & \cmark & \xmark & \xmark & \cmark & \xmark \\
			
			Zafeiropoulos et al. \cite{Zafeiropoulos2022RLAutoscaling} &
			\xmark & \cmark & \cmark & \cmark & \cmark & \xmark & \xmark & \cmark & \xmark \\
			
			Khaleq and Ra \cite{khaleq2023intelligent} &
			\xmark & \xmark & \xmark & \xmark & \xmark & \xmark & \xmark & \cmark & \xmark \\
			
			Benedetti et al. \cite{benedetti2025qlearning} &
			\xmark & \xmark & \xmark & \cmark & \xmark & \xmark & \xmark & \cmark & \xmark \\
			
			Mampage et al. \cite{mampage2025deep} &
			\cmark & \cmark & \cmark & \cmark & \cmark & \xmark & \xmark & \cmark & \cmark \\
			
			Li et al. \cite{li2025energy} &
			\cmark & \xmark & \xmark & \xmark & \xmark & \xmark & \xmark & \cmark & \xmark \\
			
			Abdullah et al. \cite{abdullah2025dynamic} &
			\xmark & \xmark & \xmark & \xmark & \cmark & \xmark & \xmark & \cmark & \xmark \\
			
			Ship et al. \cite{ship2024optimizing} &
			\cmark & \cmark & \cmark & \cmark & \xmark & \cmark & \xmark & \cmark & \cmark \\
			
			Zafarzade and Ataie \cite{zafarzade2025capacity} &
			\cmark & \cmark & \xmark & \cmark & \xmark & \xmark & \xmark & \cmark & \xmark \\
			
			Wu et al. \cite{wu2025cloudnativesim} &
			\xmark & \cmark & \cmark & \xmark & \xmark & \cmark & \xmark & \cmark & \cmark \\
			
			Fé et al. \cite{fe2022performance} &
			\cmark & \cmark & \xmark & \xmark & \xmark & \cmark & \cmark & \cmark & \xmark \\
			
			Gias et al. \cite{gias2019atom} &
			\cmark & \xmark & \xmark & \xmark & \cmark & \xmark & \xmark & \cmark & \cmark \\
			
			Luan and Shen \cite{luan2024minimize} &
			\cmark & \xmark & \xmark & \xmark & \xmark & \xmark & \xmark & \xmark & \xmark \\
			
			Gao et al. \cite{gao2024performance} &
			\cmark & \cmark & \xmark & \xmark & \xmark & \xmark & \xmark & \xmark & \xmark \\
			
			Lim et al. \cite{lim2026sinas} &
			\cmark & \xmark & \cmark & \xmark & \cmark & \xmark & \xmark & \cmark & \xmark \\
			
			Ataie et al. \cite{ataie2022modeling} &
			\cmark & \cmark & \xmark & \xmark & \xmark & \cmark & \xmark & \xmark & \xmark \\
			
			Merkouche et al. \cite{merkouche2023petri} &
			\cmark & \xmark & \xmark & \xmark & \xmark & \xmark & \xmark & \xmark & \xmark \\
			
			Jawaddi et al. \cite{jawaddi2025energy} &
			\cmark & \xmark & \cmark & \cmark & \xmark & \xmark & \xmark & \cmark & \xmark \\

			\textbf{Proposed Framework} &
			\textbf{\cmark} &
			\textbf{\cmark} &
			\textbf{\cmark} &
			\textbf{\cmark} &
			\textbf{\cmark} &
			\textbf{\cmark} &
			\textbf{\cmark} &
			\textbf{\cmark} &
			\textbf{\cmark}
			\\
			
			\bottomrule
		\end{tabular}
	}
\end{table*}

\section{Background}
\label{sec:background_system}

A QN provides a mathematical framework for representing systems in which
requests compete for processing resources. A QN consists of
interconnected service stations through which requests may arrive,
receive service, wait when processing capacity is occupied, and move
between stations according to the system structure. Each station can
be characterized by parameters such as the arrival process,
service-time distribution, number of servers, and scheduling policy.
A QN enables performance measures such as throughput, response time,
utilization, and system population to be analyzed under different
workload and capacity conditions. In this study, a single-station
$M/M/\infty$ queueing model is used as the analytical baseline for
SPR, representing an idealized system in which each incoming request
can immediately obtain processing capacity without queueing
\citep{Jackson1957Networks,Jackson1963Jobshop}.

A PN provides a state-based formalism for representing the dynamic
behavior of systems with concurrent, synchronized, and conditional
activities. A PN consists of places, transitions, tokens, and directed
arcs connecting places and transitions. Places represent system
conditions or resource states, while tokens located in places represent
system entities or quantities associated with these states, such as
requests, available capacity, or concurrency permits. The distribution
of tokens across places, referred to as the marking, defines the current
state of the modeled system. Transitions represent events whose firing
changes the system state by consuming tokens from input places and
producing tokens in output places when enabling conditions are satisfied.
This structure makes PN suitable for representing concurrency,
synchronization, resource sharing, and conditional behavior in dynamic
systems \citep{Murata1989PetriNets}.

QPN combines the state-based representation of PN with queueing and service behavior \citep{Bause1993QPN}. In particular, queueing places allow tokens to wait for and receive service according to a specified scheduling strategy and service-time distribution. QPN can therefore represent both the logical evolution of the system state and performance-related effects such as queueing and processing delays.

In this study, these formalisms provide the conceptual and analytical basis for modeling the considered autoscaling mechanisms. QN is used to establish the analytical baseline, while PN concepts provide the state-based mechanisms required to represent conditional activation and resource-state changes. QPN incorporates these mechanisms with queueing and service dynamics, providing the basis for modeling the dynamic behavior of the autoscaling strategies developed in Section~\ref{sec:Proposed}.


\section{System Description}
\label{sec:desc_system}

\begin{figure*}[!t] 
	\centering \begin{minipage}[t]{0.32\textwidth} \centering \includegraphics[width=\linewidth]{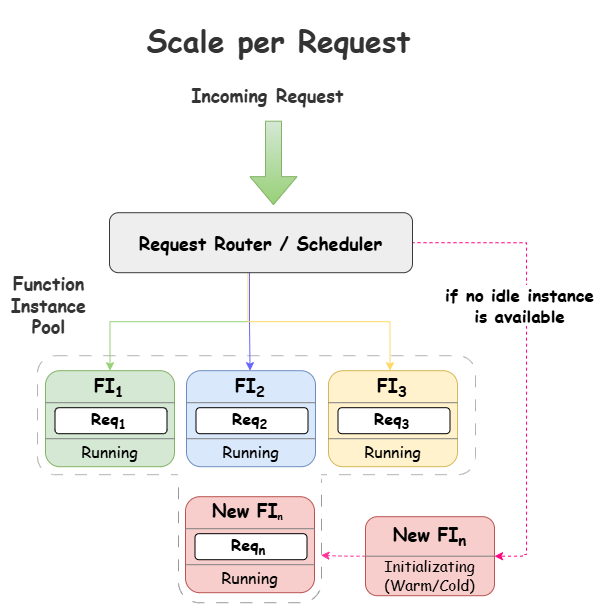} \vspace{2mm} \textbf{(a) SPR} \end{minipage} \hfill \begin{minipage}[t]{0.32\textwidth} \centering \includegraphics[width=\linewidth]{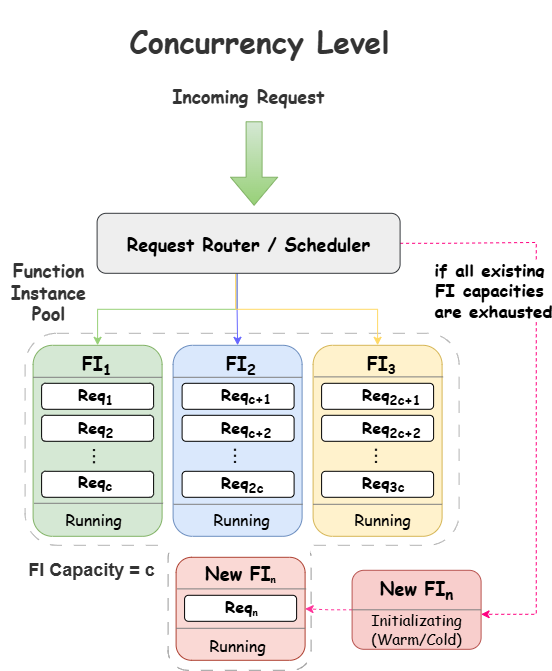} \vspace{2mm} \textbf{(b) CLS} \end{minipage} \hfill \begin{minipage}[t]{0.32\textwidth} \centering \includegraphics[width=\linewidth]{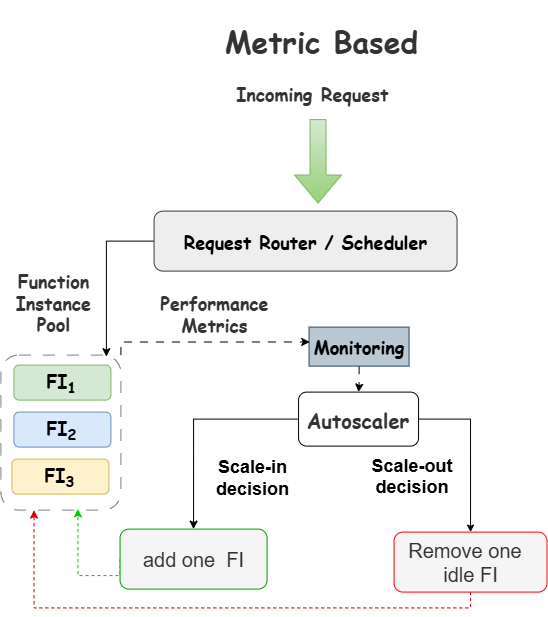} \vspace{2mm} \textbf{(c) MBS} \end{minipage} \caption{Conceptual illustrations of the autoscaling approaches considered in this study: (a) SPR, (b) CLS, and (c) MBS.} \label{fig:scaling_models} 
\end{figure*}

Autoscaling enables computing capacity to be adjusted dynamically
according to workload conditions, reducing the need for direct
intervention in resource provisioning. Autoscaling mechanisms can be
broadly categorized as horizontal or vertical. Horizontal scaling,
also referred to as scale-out/scale-in, adjusts capacity by increasing
or decreasing the number of active FIs. In contrast, vertical scaling,
or scale-up/scale-down, modifies the computational resources, such as CPU or memory, allocated
to an existing FI.

The autoscaling strategies evaluated in this research primarily follow
the horizontal scaling paradigm, since capacity adaptation is achieved
through the activation or deactivation of FIs. Three representative
approaches are considered here: SPR, CLS, and MBS. The descriptions in
this section provide a conceptual view of these autoscaling approaches,
while their corresponding QPN implementations and evaluation settings
are presented in Section~\ref{sec:Proposed}.

\paragraph{SPR.} In SPR \citep{Mahmoudi2022Serverless}, processing capacity is adapted on a per-request basis. As illustrated in Fig.~\ref{fig:scaling_models}(a), incoming requests are directed to a request router/scheduler, which assigns each request to an available FI. 
Each FI can process at most one request at a time. Therefore, if an idle FI is available, it can be reused to process a new request.

If no idle FI is available when a request arrives, an additional FI is activated to provide the required processing capacity. After completing the assigned task, the FI is retained in a warm state for a predefined idle period and becomes available for subsequent requests during this period \citep{ACM_TOCS_2025}. If no new request arrives within the defined retention interval, the FI is released and its allocated resources are removed from the active capacity pool. Thus, in SPR, processing capacity is adapted directly according to request arrivals and FI availability, rather than being triggered by a concurrency limit or a monitored system metric \citep{Mahmoudi2022Serverless, Anselmi2025SPR}.

\paragraph{CLS.}
In CLS \citep{GoogleCloudRunConcurrency}, each FI is configured with a
predefined concurrency capacity, denoted by $c$, which determines the
maximum number of requests that can be processed simultaneously
\citep{IEEE_Access_2024}. As illustrated in
Fig.~\ref{fig:scaling_models}(b), incoming requests are directed to a
request router/scheduler and assigned to an FI with available
concurrency capacity.

Requests continue to be assigned to an FI until its concurrency
capacity is fully utilized. When the available capacity of all active
FIs is exhausted and additional requests require service, another FI
is activated, contributing an additional concurrency capacity of $c$
\citep{IEEE_TCC_2025}. This process continues progressively as the
workload increases. When a request completes execution, the
corresponding concurrency capacity of the host FI becomes available
again and can be reused by newly arriving or waiting requests before
additional FIs are activated. After becoming idle, an FI remains
available for a predefined retention period. If no new requests are
assigned to the FI during this period, the FI is released and the
number of active FIs is reduced. Thus, CLS performs scale-out when
the available concurrency capacity is exhausted and scale-in when
excess FIs remain idle beyond the retention period, while reusing
available capacity in active FIs whenever possible
\citep{GoogleCloudRunConcurrency, IEEE_TCC_2025}.

\paragraph{MBS.}
In MBS \citep{AWSLambdaAutoScaling, Mahmoudi2023MetricBased}, scaling decisions
are based on monitored system metrics that reflect the current workload
conditions. Several metrics can be considered for this purpose,
including response time, utilization, queue length, throughput, and
cost-related indicators. In this study, utilization is used as the
primary scaling metric. As illustrated in
Fig.~\ref{fig:scaling_models}(c), incoming requests are distributed
among the active FIs, while their utilization is monitored. The
observed utilization is provided to the autoscaler, which compares it
with predefined scaling thresholds \citep{IEEE_TSC_2026}.

When utilization exceeds the upper threshold, the autoscaler activates
an additional FI to increase the available processing capacity.
Conversely, when utilization falls below the lower threshold, excess
FIs can be deactivated as the workload decreases. The conceptual view
in Fig.~\ref{fig:scaling_models}(c) also includes a warm state,
representing the temporary retention of an idle FI before it is
released in practical serverless environments
\citep{IEEE_TSC_2026}. Overall, MBS adjusts the number of active FIs
according to monitored utilization, providing scale-out under
increasing demand and scale-in as workload demand decreases.

Table~\ref{tbl:comp-approaches} summarizes the main characteristics of
the scaling approaches considered in this paper.

\begin{table*}[t]
	\caption{Comparison of the SPR, CLS, and MBS autoscaling approaches.}
	\centering
	\label{tbl:comp-approaches}
	\resizebox{\textwidth}{!}{%
		\begin{tabular}{@{} p{3.5cm} p{4.0cm} p{4.5cm} p{4.5cm} @{}}
			\toprule
			\textbf{Feature} &
			\textbf{SPR} &
			\textbf{CLS} &
			\textbf{MBS} \\
			\midrule
			
			Scaling Type &
			Horizontal &
			Horizontal, Vertical &
			Horizontal \\
			
			Model Structure &
			One concurrent request per FI &
			Shared FIs with concurrency limits &
			Utilization-driven FI activation \\
			
			Modeling Tool &
			JSIMgraph, MATLAB &
			JSIMgraph, Python &
			JSIMgraph, Python \\
			
			Execution Policy &
			Independent per-request processing &
			Concurrency-limited shared processing &
			Threshold-triggered scaling with routing \\
			
			Scaling Trigger/Metric &
			Request arrival &
			Concurrency limit &
			Utilization \\
			
			Allocated Resources &
			Dedicated FI capacity per request &
			Shared FI capacity subject to concurrency limits &
			FI count adapted according to utilization \\
			
			\bottomrule
		\end{tabular}%
	}
\end{table*}

\section{The Proposed Model}
\label{sec:Proposed}

This section presents the models developed for the SPR, CLS, and
MBS autoscaling strategies. The operational behavior described in
Section~\ref{sec:desc_system} is represented using the modeling
formalisms introduced in Section~\ref{sec:background_system}. For each
strategy, the model components and request-flow mechanism are described
according to their corresponding formalism. The performance metrics
used for the comparative evaluation are then introduced.

\begin{figure}[pos=H]
	\centering
	\includegraphics[
	width=\columnwidth,
	keepaspectratio
	]{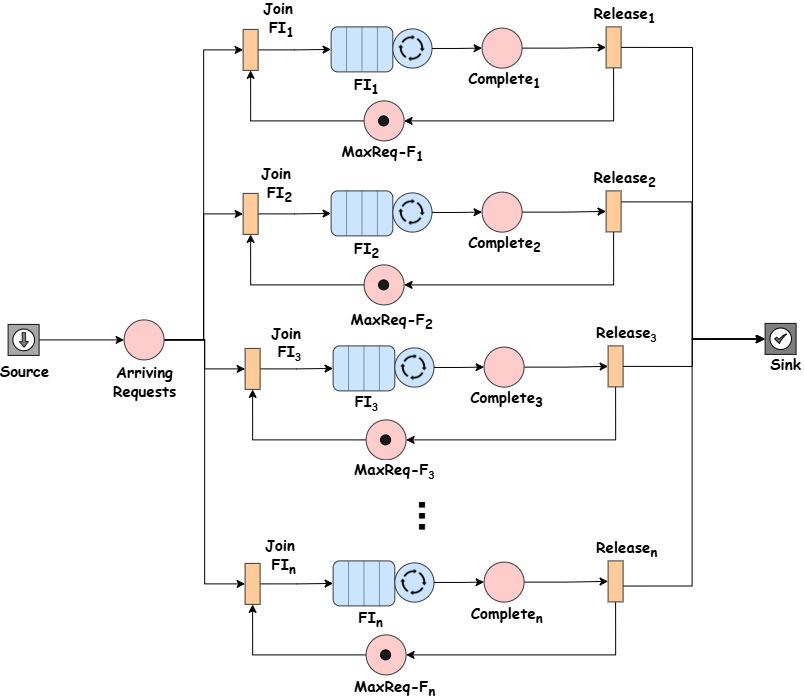}
	\caption{QPN representation of the SPR architecture.}
	\label{fig:spr_qpn}
\end{figure}

\subsection{Analytical and QPN-Based Autoscaling Models}
\label{subsec:model_architectures}

The three autoscaling strategies are represented using complementary
modeling formalisms tailored to their operational mechanics. In
Section~\ref{subsubsec:spr_model}, the SPR strategy is modeled using
a combination of an analytical QN and a detailed QPN representation
to capture explicit request flows and instance-level capacity
management. In Section~\ref{subsubsec:cl_model}, the CLS strategy is
represented using QPN to capture dynamic concurrency-control behavior
and the state-dependent activation of FI capacity. Finally,
Section~\ref{subsubsec:aspm_model} presents the QPN-based framework
for the MBS strategy, where metric-driven scaling decisions and
request dispatching algorithms are integrated within a central control
component.

\subsubsection{SPR Model}
\label{subsubsec:spr_model}

For SPR, an $M/M/\infty$ queue is first used as an analytical baseline.
The model represents an idealized processing system in which each
arriving request can immediately obtain processing capacity and
therefore does not wait because of a shortage of available FIs. To
represent the internal request flow and FI-capacity control explicitly,
SPR is also modeled using QPN, as shown in
Fig.~\ref{fig:spr_qpn}.

In Fig.~\ref{fig:spr_qpn}, the \texttt{Source} generates incoming
requests, which are represented by request tokens in the QPN model.
These tokens enter the \texttt{Arriving Requests} place, where they
wait until an FI branch with available processing capacity can admit
them. The model contains a set of parallel FI branches, represented
by \texttt{FI}\(_1\), \texttt{FI}\(_2\), \texttt{FI}\(_3\),
$\ldots$, \texttt{FI}\(_n\). Each branch has a corresponding
\texttt{MaxReq-F}\(_i\) place containing a capacity token that
represents the availability of the associated \texttt{FI}\(_i\).
Thus, the presence of a token in \texttt{MaxReq-F}\(_i\) indicates
that the corresponding FI is available to admit a request.

For each branch, the corresponding \texttt{Join FI}\(_i\) component
is modeled as a \emph{transition}. It has two required inputs: a
request token from the \texttt{Arriving Requests} place and a
capacity token from \texttt{MaxReq-F}\(_i\). Therefore,
\texttt{Join FI}\(_i\) becomes enabled only when both a waiting
request and available FI capacity are present. When the transition
fires, one request token is removed from
\texttt{Arriving Requests} and forwarded to the corresponding
\texttt{FI}\(_i\) queue, while the associated capacity token is
removed from \texttt{MaxReq-F}\(_i\). Removing this capacity token
prevents the same FI from admitting another request while its current
request is being processed.

For example, in the first branch, \texttt{Join FI}\(_1\) requires a
request token from \texttt{Arriving Requests} and the capacity token
stored in \texttt{MaxReq-F}\(_1\). Once both tokens are available,
the transition fires and the admitted request is forwarded to
\texttt{FI}\(_1\), while the capacity token is temporarily removed
from \texttt{MaxReq-F}\(_1\).

Each \texttt{FI}\(_i\) component is modeled as a
\emph{queueing/service station} that represents request execution by
the corr.esponding function instance. After service is completed, the
request token leaves \texttt{FI}\(_i\) and enters the corresponding
\texttt{Complete}\(_i\) \emph{place}. This place represents the
state in which processing has finished but the completed request has
not yet departed from the model

The corresponding \texttt{Release}\(_i\) component is modeled as a
\emph{transition} that performs both request departure and capacity
restoration. When \texttt{Release}\(_i\) fires, the completed request
token is forwarded from \texttt{Complete}\(_i\) to the common
\texttt{Sink}, which represents the departure of completed requests
from the system. At the same time, the capacity token associated with
\texttt{FI}\(_i\) is returned to \texttt{MaxReq-F}\(_i\). The return
of this token indicates that the corresponding \texttt{FI}\(_i\) is
again available to admit and process another request.

The same sequence of QPN components is replicated across all FI
branches, including \texttt{FI}\(_1\), \texttt{FI}\(_2\),
\texttt{FI}\(_3\), and the subsequent branches up to
\texttt{FI}\(_n\). Accordingly, each branch contains its
corresponding \texttt{MaxReq-F}\(_i\), \texttt{Join FI}\(_i\),
\texttt{FI}\(_i\), \texttt{Complete}\(_i\), and
\texttt{Release}\(_i\) components, where $i=1,\ldots,n$.
All FI branches receive request tokens from the common
\texttt{Arriving Requests} place, while their release transitions
forward completed request tokens to the same \texttt{Sink}.
Therefore, multiple requests can be processed concurrently through
separate FI branches whenever their corresponding capacity tokens
are available. In this way, the QPN model represents the SPR
mechanism through request tokens, FI-capacity tokens, admission
transitions, parallel service stations, completion places, and
release transitions.

\subsubsection{CLS Model}
\label{subsubsec:cl_model}

\begin{figure}[pos=H]
	\centering
	\includegraphics[
	width=\columnwidth,
	height=0.35\textheight,
	keepaspectratio
	]{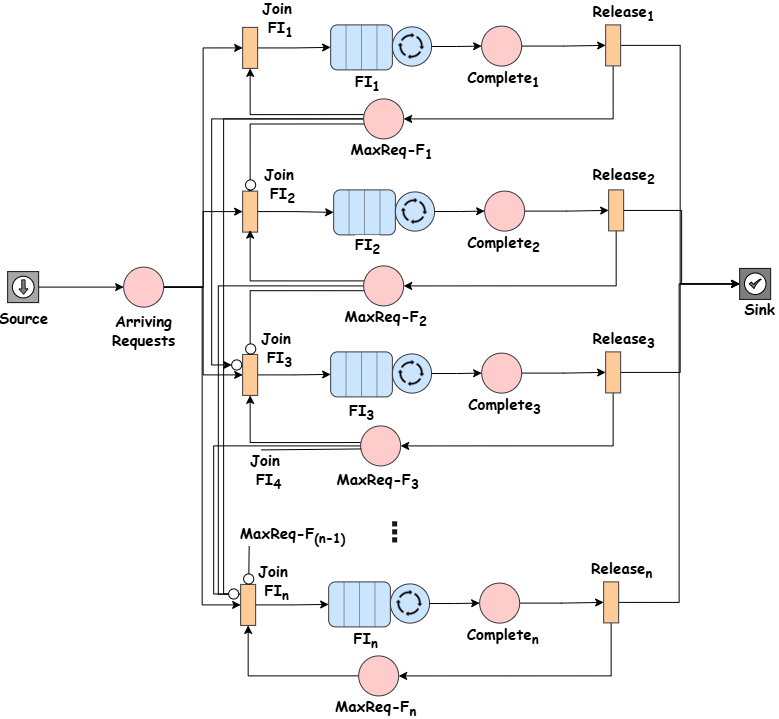}
	\caption{QPN representation of the CLS architecture with
		concurrency control and conditional FI activation.}
	\label{fig:cl_qpn}
\end{figure}

The CLS strategy is represented using QPN to capture
concurrency-limited request admission and the state-dependent
activation of additional FI capacity. The corresponding model is
shown in Fig.~\ref{fig:cl_qpn}.
The QPN representation explicitly describes how concurrency tokens
control request admission and how subsequent FI branches become
available for processing. 

The corresponding model is shown in
Fig.~\ref{fig:cl_qpn}. Requests generated by the
\texttt{Source} enter the \texttt{Arriving Requests} place, where
they wait until the conditions required for admission to an FI are
satisfied. For each FI branch, from \texttt{FI}\(_1\) to
\texttt{FI}\(_n\), a corresponding \texttt{MaxReq-F} place contains
a predefined number of concurrency tokens, where each token represents
one unit of concurrency capacity of the associated FI. Therefore, the
initial number of tokens in \texttt{MaxReq-F}\(_1\),
\texttt{MaxReq-F}\(_2\), \texttt{MaxReq-F}\(_3\), \ldots,
\texttt{MaxReq-F}\(_n\) defines the maximum number of requests that
\texttt{FI}\(_1\), \texttt{FI}\(_2\), \texttt{FI}\(_3\), \ldots,
\texttt{FI}\(_n\) can process concurrently, respectively. The
\texttt{Join FI}\(_1\), \texttt{Join FI}\(_2\),
\texttt{Join FI}\(_3\), \ldots, \texttt{Join FI}\(_n\) components are
modeled as transitions that control the admission of requests from
the \texttt{Arriving Requests} place to the corresponding FI
processing components. Inhibitor arcs impose the additional conditions
required for admitting requests to subsequent FI branches.

The \texttt{Join FI}\(_1\) component is an admission
\emph{transition}. It becomes enabled when at least one request token
is waiting in \texttt{Arriving Requests} and at least one concurrency
token is available in \texttt{MaxReq-F}\(_1\). When the transition
fires, one request token is admitted to the \texttt{FI}\(_1\)
service station and one concurrency token is removed from
\texttt{MaxReq-F}\(_1\). The consumed concurrency token remains
unavailable while the corresponding request is being processed.

The \texttt{FI}\(_1\) component is modeled as a
\emph{queueing/service station}. Multiple requests can therefore be
admitted to this station as long as concurrency tokens remain in
\texttt{MaxReq-F}\(_1\). Once all tokens in
\texttt{MaxReq-F}\(_1\) are consumed, \texttt{FI}\(_1\) reaches its
configured concurrency limit and cannot admit additional requests
until some of its concurrency capacity becomes available again.

The admission transitions of subsequent FI branches are additionally
controlled by \emph{inhibitor arcs}. In particular,
\texttt{Join FI}\(_2\) is inhibited while concurrency capacity remains
available in \texttt{MaxReq-F}\(_1\). When
\texttt{MaxReq-F}\(_1\) becomes empty, the concurrency capacity of
\texttt{FI}\(_1\) is fully utilized. Under this condition, if another
request is waiting in \texttt{Arriving Requests} and a concurrency
token is available in \texttt{MaxReq-F}\(_2\),
\texttt{Join FI}\(_2\) can admit the request to
\texttt{FI}\(_2\).

The same admission logic is extended to the remaining FI branches.
Thus, \texttt{Join FI}\(_3\) becomes eligible only after the required
preceding FI capacities have been exhausted. The same structure
continues through the intermediate FI branches, including
\texttt{Join FI}\(_{n-1}\), up to \texttt{Join FI}\(_n\). In this
way, the inhibitor-arc structure enforces sequential utilization of
the available FI concurrency capacity.

After service at any FI branch is completed, the corresponding request
token moves from the FI service station of that branch to its
associated \texttt{Complete} place, such as
\texttt{Complete}\(_1\), \texttt{Complete}\(_2\),
\texttt{Complete}\(_3\), \ldots, \texttt{Complete}\(_n\). The
associated \texttt{Release} transition of that branch, such as
\texttt{Release}\(_1\), \texttt{Release}\(_2\),
\texttt{Release}\(_3\), \ldots, \texttt{Release}\(_n\), then forwards
the completed request token to the common \texttt{Sink}, which
represents the departure of completed requests from the system. At the
same time, the consumed concurrency token is returned to the
corresponding \texttt{MaxReq-F} place of that branch. The returned
token indicates that one unit of concurrency capacity has become
available again in the corresponding FI.

The return of concurrency tokens also changes the inhibitor conditions
dynamically. For example, if \texttt{FI}\(_2\) has become available
for admission because all tokens in \texttt{MaxReq-F}\(_1\) were
consumed, the return of a token to \texttt{MaxReq-F}\(_1\) indicates
that \texttt{FI}\(_1\) again has available concurrency capacity. The
corresponding inhibitor condition therefore prevents further admission
to subsequent FI branches while capacity remains available in the
preceding branch. Waiting or newly arriving requests are consequently
directed again to the earliest FI with available concurrency capacity.

\begin{figure}[pos=H]
	\centering
	\includegraphics[
	width=\columnwidth,
	height=0.35\textheight,
	keepaspectratio
	]{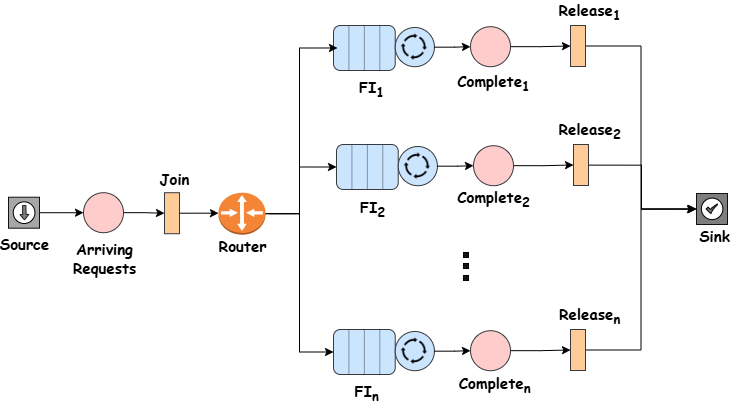}
	\caption{QPN-based representation of the MBS Sarchitecture.}
	\label{fig:aspm_arch}
\end{figure}

Thus, the CLS model explicitly combines a request waiting place, request
tokens, concurrency-capacity places, concurrency tokens, admission
transitions, FI service stations, completion places, release
transitions, inhibitor arcs, and a common sink. Together, these QPN
elements represent the sequential and state-dependent concurrency
behavior of CLS.

\subsubsection{MBS Model}
\label{subsubsec:aspm_model}

The MBS strategy is represented using a QPN-based architecture in
which metric-driven autoscaling and request dispatching are integrated
within the \texttt{Router}. The corresponding model is shown in
Fig.~\ref{fig:aspm_arch}.

In Fig.~\ref{fig:aspm_arch}, the \texttt{Source} generates incoming
requests, which are represented by request tokens in the QPN model.
These tokens first enter the \texttt{Arriving Requests} place, where
they wait before entering the control and processing part of the
model. The \texttt{Join} component is modeled as a
\emph{transition} that admits waiting request tokens from
\texttt{Arriving Requests} and forwards them to the
\texttt{Router}, where the routing decision is performed.

The \texttt{Router} acts as the central control component of the
MBS model and performs both metric-driven autoscaling and request
dispatching. During system operation, the \texttt{Router} evaluates
the selected scaling metric of the active FI pool, such as utilization.
The observed metric value is compared with predefined upper and lower
thresholds to determine whether the currently available processing
capacity should be increased, maintained, or reduced.

When the monitored metric reaches or exceeds the upper threshold, the
\texttt{Router} performs a scale-out operation by making an additional
FI available for request processing. Conversely, when the monitored
metric reaches or falls below the lower threshold, the
\texttt{Router} initiates the scale-in procedure. An FI that becomes
idle is not removed immediately. Instead, it remains available for a
predefined observation period and is removed only if no new request is
assigned to it during that period. Therefore, the number of active FIs
is dynamically adjusted according to the monitored workload
conditions.

The \texttt{Router} also dispatches each admitted request token among
the currently active FI branches according to the configured routing
policy. Consequently, autoscaling and request dispatching are
integrated within the same control component: the autoscaling logic
determines the processing capacity currently available, while the
routing logic determines how incoming requests are distributed over
that capacity.

Three request-dispatching algorithms are implemented and evaluated
within the \texttt{Router}: random, round-robin, and probabilistic
routing. Under random routing, one of the currently active FIs is
selected randomly for each request. Under round-robin routing,
successive request tokens are distributed cyclically among the active
FI branches. Under probabilistic routing, requests are dispatched
according to predefined routing probabilities. Thus, the selected
routing policy determines how the incoming workload is distributed
among the FIs that are active at the time of the routing decision.

The \texttt{FI}\(_1\), \texttt{FI}\(_2\), \ldots,
\texttt{FI}\(_n\) components represent the processing elements of the
model. Each request token leaving the \texttt{Router} is forwarded to
one of the active FI service stations according to the selected
dispatching algorithm. As the autoscaling condition changes, the set
of active FIs considered by the \texttt{Router} for subsequent request
dispatching is updated accordingly.

After processing is completed at an FI, the corresponding request
token moves to its associated completion place. These places are
represented by \texttt{Complete}\(_1\), \texttt{Complete}\(_2\),
\ldots, \texttt{Complete}\(_n\). The corresponding
\texttt{Release}\(_1\), \texttt{Release}\(_2\), \ldots,
\texttt{Release}\(_n\) transitions then forward the completed request
tokens to the common \texttt{Sink}, which represents their departure
from the system.

Consequently, the MBS model consists of a sequence of explicit QPN
and control components. The \texttt{Source} generates request tokens,
the \texttt{Arriving Requests} place stores waiting requests, and the
\texttt{Join} transition admits them to the \texttt{Router}. The
\texttt{Router} evaluates the selected scaling metric, performs
scale-out or scale-in when the corresponding threshold condition is
satisfied, and dispatches admitted requests among the currently active
FI branches. The \texttt{FI}\(_1\) through \texttt{FI}\(_n\) service
stations perform request processing, their corresponding
\texttt{Complete} places represent completed processing, the
associated \texttt{Release} transitions complete the request path,
and the common \texttt{Sink} removes completed request tokens from
the system.

This structure allows the autoscaling mechanism and the
request-dispatching algorithm to be investigated jointly within the
same MBS model. The monitored metric and its scaling thresholds
determine the number of active FIs, whereas the selected routing
algorithm determines how incoming requests are distributed among
those active FIs.

\subsection{Evaluation Metrics}
\label{subsec:kpis}
To rigorously evaluate and compare the performance of the considered autoscaling strategies, four system-level performance measures are defined: throughput, mean system response time, in-system population, and system power. These metrics provide a comprehensive perspective on dynamic capacity adaptation and capture the fundamental trade-offs between system responsiveness, usage level, and overall stability under varying workload conditions. The formal mathematical definitions of these metrics are detailed below.

\paragraph{Throughput ($X_{sys}$)}
represents the average rate at which requests are successfully processed and departed from the system per time unit. The overall system throughput is defined as:
\begin{equation}
	\label{eq:x}
	X_{sys} = \lambda_{in}
\end{equation}
where $\lambda_{in}$ denotes the aggregate mean arrival rate of incoming requests to the system.

\paragraph{Mean system response time ($E[T_R^{sys}]$)}
is the average of the total residence time experienced by each request across all visited processing stations. Therefore, it can be computed as the sum of the station-level mean response times ($E[T_R^i]$), which account for both queueing delay ($E[T_Q^i]$) and service time ($E[S^i]$) at station $i$:
\begin{equation}
	\label{eq:RT}
	E[T_R^{sys}] = \sum_{i=1}^{n} E[T_R^i] = \sum_{i=1}^{n} \left( E[T_Q^i] + E[S^i] \right)
\end{equation}
where $n$ represents the total number of queueing stations in the system.

\paragraph{Utilization ($U^i$)}
measures the proportion of time a station's servers are busy, which mathematically corresponds to the mean number of requests currently being served at station $i$. For a station with $M_i$ servers, effective arrival rate $\lambda_i$, and mean service time $E[S^i]$, utilization is calculated as:
\begin{equation}
	\label{eq:Util}
	U^i = \frac{\lambda_i \cdot E[S^i]}{M_i}
\end{equation}
\paragraph{In-system population ($E[N_{req}^{sys}]$)}
quantifies the average of the total number of requests presented across the entire system. At the station level, the expected population ($E[N_{req}^i]$) is the sum of the requests waiting in the queue ($E[N_Q^i]$) of station $i$, and those actively being served ($U^i$) at that station. Therefore, the expected in-system population is expressed as:
\begin{equation}
	\label{eq:n_req}
	E[N_{req}^{sys}] = \sum_{i=1}^{n} E[N_{req}^i] = \sum_{i=1}^{n} \left( E[N_Q^i] + U^i \right)
\end{equation}
\paragraph{System power ($P_{sys}$)}
characterizes the optimal operational efficiency of the system by evaluating the speed--throughput trade-off. It is defined as the ratio of overall system throughput to mean system response time:
\begin{equation}
	\label{eq:power}
	P_{sys} = \frac{X_{sys}}{E[T_R^{sys}]}
\end{equation}
A higher system power value indicates an optimal operating state where the system maximizes throughput while keeping latency low.



\section{Performance Evaluation Results}
\label{sec:performance_results}

This section presents the performance evaluation of SPR, CLS, and MBS.
Section~\ref{subsec:common_setup} first defines the common experimental
settings used throughout the simulation-based evaluations. The results
for SPR, CLS, and MBS are then presented, followed by a comparative
evaluation of the three strategies.

\subsection{Common Experimental Setup}
\label{subsec:common_setup}

Unless otherwise stated, the simulation-based evaluations use the same
$H_2$ request-arrival process. The workload is characterized by a
mixing probability of $p=0.2$ and a squared coefficient of variation
of the inter-arrival time of $C_a^2 \approx 2.125$, corresponding to
$C_a \approx 1.457$. This level of variability reflects the bursty
nature of the workload.

In the JMT \emph{What-if analysis}, $\lambda$ denotes the mean arrival
rate of the overall $H_2$ workload and is used as the workload-control
parameter. When $\lambda$ is varied, JMT adjusts the mean inter-arrival
time while preserving the mixing probability $p$ and the coefficient
of variation of the arrival process; the underlying phase rates are
therefore recalculated accordingly
\cite{Serazzi2024Variability,JMTManual,Bertoli2009JMT}.

Accordingly, the workload levels reported in the following experiments
	represent mean arrival rates rather than constant instantaneous ones. Because the $H_2$ arrival process generates inter-arrival times
	probabilistically from its two phases, requests may occasionally arrive
	in closely spaced groups, creating short-term bursts even when the mean
	arrival rate remains unchanged.

\subsection{SPR Performance Evaluation}
\label{subsec:spr_results}

The SPR strategy is evaluated from both idealized and practical
perspectives. First, an unlimited-capacity analytical baseline is
validated against simulation results. The analysis is then extended
to a capacity-limited implementation to examine performance
degradation near saturation and the effects of finite resource
availability.
\subsubsection{Analytical Baseline and Model Validation}
\label{subsubsec:spr_validation}

To establish an ideal analytical baseline for the SPR strategy, an
$M/M/\infty$ model is considered, in which requests arrive according
to a Poisson process with rate $\lambda_{in}$, service times are
exponentially distributed with rate $\mu$, and each arriving request
can immediately obtain processing capacity. Under this idealized
assumption, no queueing delay is introduced.

The analytical calculations are carried out in MATLAB, and the resulting
values are compared with unlimited-capacity simulation results to validate
the model implementation. For each arrival rate, five independent
simulation runs are performed, and the resulting performance measures
are averaged.

For the unlimited-capacity model, every arriving request immediately
enters service; therefore, system throughput follows the input rate,
i.e., $X_{sys}=\lambda_{in}$, as defined in
Equation~\ref{eq:x}. Since the queueing delay is zero, the mean system
response time is determined only by the mean service time. With the
service rate set to $\mu=3$ req/s,

\begin{equation}
	\label{eq:response_time}
	E[T_R^{sys}]
	=
	E[S]
	=
	\frac{1}{\mu}
	\approx 0.33~\text{seconds}.
\end{equation}

The corresponding mean system population is obtained from Little's
Law as

\begin{equation}
	\label{eq:little_spr}
	E[N_{req}^{sys}]
	=
	\lambda_{in}E[T_R^{sys}].
\end{equation}

The simulation results closely reproduce these analytical
relationships. System throughput follows the input arrival rate over
the evaluated workload range, as expected from the $M/M/\infty$
formulation, with a maximum relative deviation of approximately
$0.3\%$ between the simulated and analytical values. The simulated
response time remains close to the analytical value of $0.33$ seconds;
relatively larger deviations at low arrival rates result from the
smaller number of completed requests available for estimating the
mean response time. The relative response-time error is approximately
$0.49\%$. The simulated system population also follows the linear
behavior predicted by Little's Law, with an analytical mean population
of approximately $5$ requests at $\lambda_{in}=15$ req/s.

Using the throughput and response-time results above in
Equation~\ref{eq:power},
the analytical system power increases linearly with the arrival rate and reaches approximately
$45$
at
$\lambda_{in}=15$ req/s.
The simulation-derived system power follows the same approximately linear increasing trend as the analytical baseline.

This initial validation considers system throughput, mean system
response time, system population, and system power. Overall, the agreement between the
analytical and simulation results validates the unlimited-capacity
SPR implementation and establishes the ideal analytical baseline used
for comparison with the capacity-limited SPR model in
Section~\ref{subsubsec:spr_capacity_limited}.

\begin{figure*}[pos=H]
	\centering
	
	\begin{subfigure}[t]{0.32\textwidth}
		\centering
		\includegraphics[
		width=\linewidth,
		height=0.22\textheight,
		keepaspectratio
		]{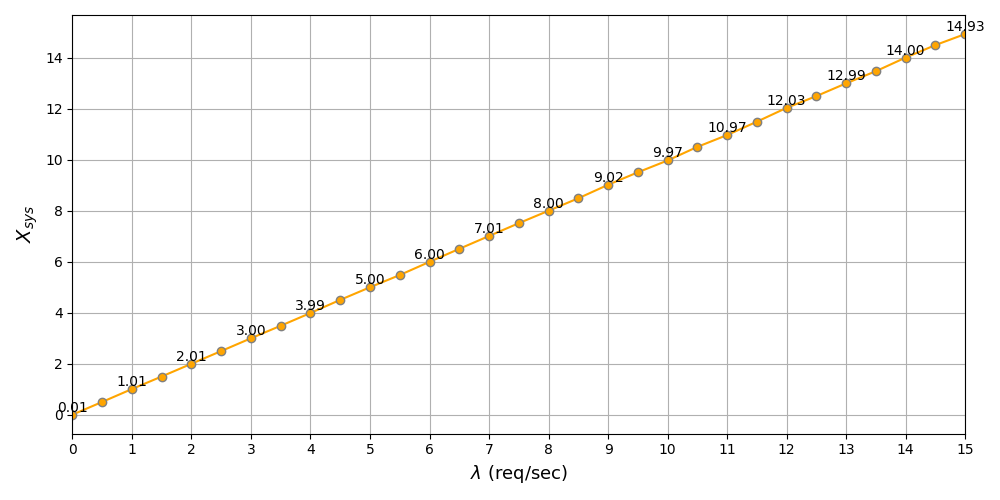}
		\caption{System throughput.}
		\label{fig:spr_capacity_throughput}
	\end{subfigure}
	\hfill
	\begin{subfigure}[t]{0.32\textwidth}
		\centering
		\includegraphics[
		width=\linewidth,
		height=0.22\textheight,
		keepaspectratio
		]{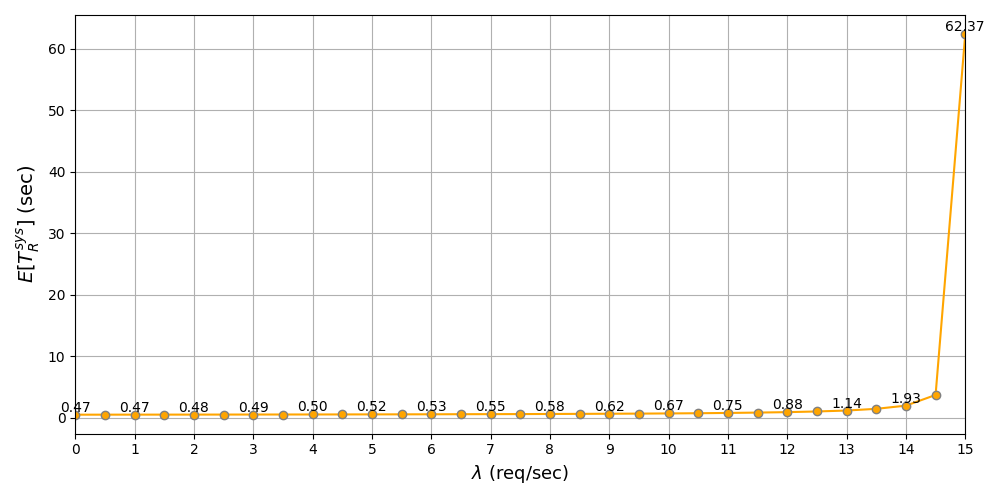}
		\caption{Mean system response time.}
		\label{fig:spr_capacity_response}
	\end{subfigure}
	\hfill
	\begin{subfigure}[t]{0.32\textwidth}
		\centering
		\includegraphics[
		width=\linewidth,
		height=0.22\textheight,
		keepaspectratio
		]{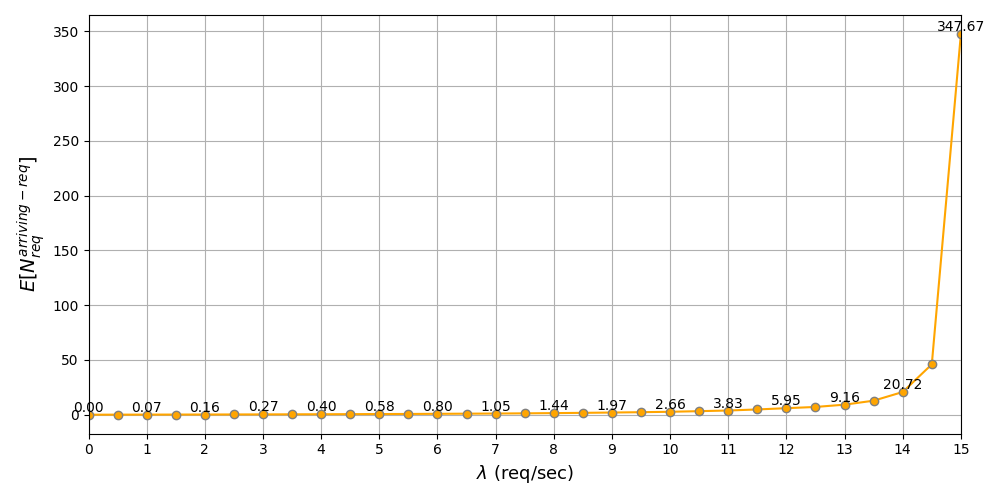}
		\caption{Mean number of requests in the
			\texttt{Arriving Requests} input place.}
		\label{fig:spr_capacity_queue}
	\end{subfigure}
	
	\vspace{0.35cm}
	
	\begin{subfigure}[t]{0.47\textwidth}
		\centering
		\includegraphics[
		width=\linewidth,
		height=0.24\textheight,
		keepaspectratio
		]{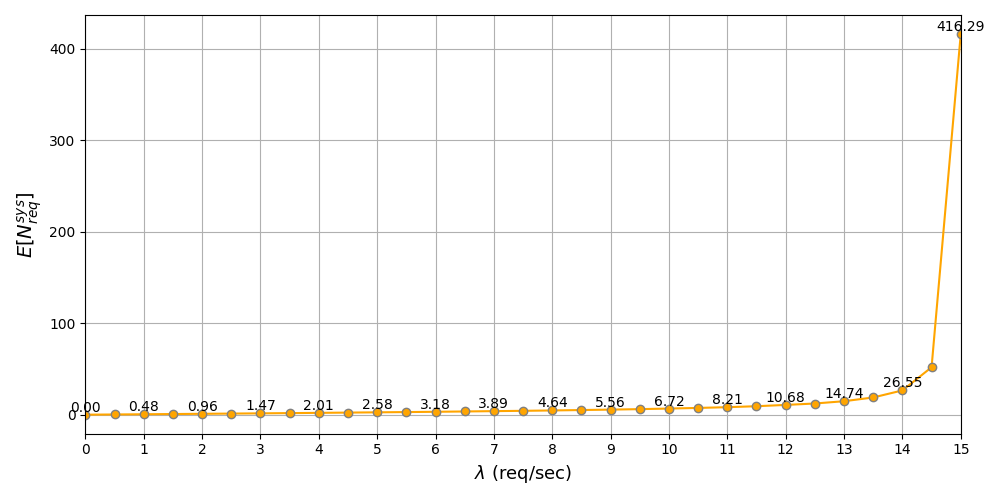}
		\caption{System population.}
		\label{fig:spr_capacity_population}
	\end{subfigure}
	\hfill
	\begin{subfigure}[t]{0.47\textwidth}
		\centering
		\includegraphics[
		width=\linewidth,
		height=0.24\textheight,
		keepaspectratio
		]{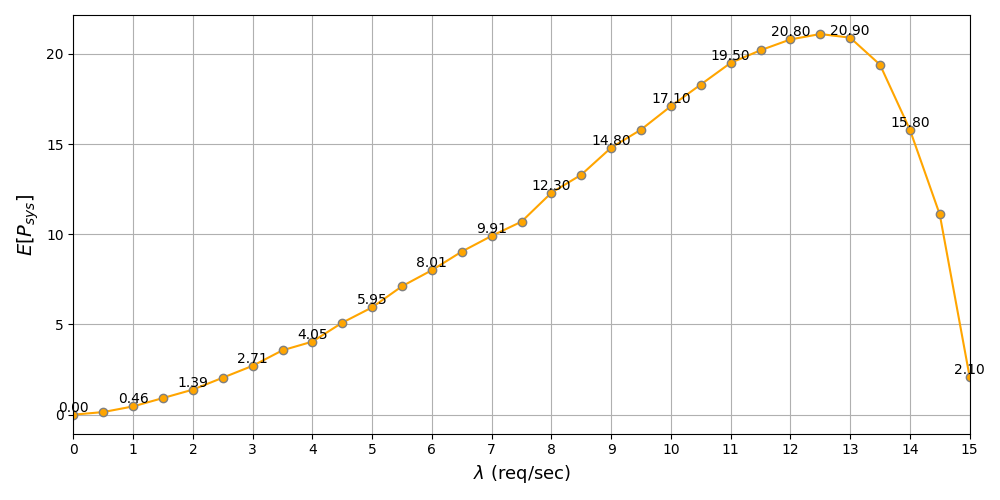}
		\caption{System power.}
		\label{fig:spr_capacity_power}
	\end{subfigure}
	
	\caption{Performance of the capacity-limited SPR model under
		increasing workload: (a) system throughput, (b) mean system
		response time, (c) mean number of requests in the
		\texttt{Arriving Requests} input place, (d) system
		population, and (e) system power.}
	\label{fig:spr_capacity_results}
\end{figure*}

\subsubsection{Capacity-Limited SPR Performance}
\label{subsubsec:spr_capacity_limited}

To evaluate SPR under finite processing capacity, the practical QPN
model described in Section~\ref{subsubsec:spr_model} is implemented
and evaluated in JMT. Unlike the unlimited-capacity baseline considered
in Section~\ref{subsubsec:spr_validation}, this model is limited to
15 simultaneously active FIs, with each FI processing at most one
request at a time. The capacity-limited model uses the common $H_2$
workload defined in Section~\ref{subsec:common_setup}. Service times
are exponentially distributed with a service rate of $\mu=3$ req/s.
The simulation is executed with up to 500,000 samples and repeated
10 times, and the reported performance measures are obtained by
averaging the corresponding independent runs.

As shown in Fig.~\ref{fig:spr_capacity_results}
(\subref{fig:spr_capacity_throughput}), system throughput increases
almost linearly with the mean arrival rate $\lambda$ and remains close
to the ideal analytical baseline throughout most of the evaluated
workload range. The mean relative error between the capacity-limited
QPN results and the analytical SPR baseline is approximately $0.54\%$.
The impact of finite capacity becomes considerably more evident in
the mean system response time, as illustrated in
Fig.~\ref{fig:spr_capacity_results}
(\subref{fig:spr_capacity_response}). Under low and moderate loads,
the response time remains relatively stable, although it is
approximately $40\text{--}50\%$ higher than the ideal
unlimited-capacity baseline.
As the arrival rate approaches the
capacity provided by the 15 parallel FIs, the response-time curve
begins to deteriorate rapidly.
Since
$\lambda$
denotes the mean arrival rate of the bursty
$H_2$
workload rather than a constant instantaneous rate, short-term
	arrival bursts may occur around a given mean workload level and
	contribute to the observed increase in delay.

A practical breaking point is observed at approximately
$\lambda=14.5$ req/s, beyond which the mean response time increases
sharply and reaches approximately $62.37$ seconds at
$\lambda=15$ req/s.

To further examine the source of the rapid response-time increase,
the mean number of requests accumulated in the
\texttt{Arriving Requests} input place is also examined
(Fig.~\ref{fig:spr_capacity_results}
(\subref{fig:spr_capacity_queue})). This quantity represents the
requests that remain in the input place before they can proceed to
the processing part of the SPR model. At low and moderate arrival
rates, the population of this place remains relatively small. As the
workload approaches the high-load region, however, requests begin to
accumulate rapidly in the input place. At $\lambda=15$ req/s, the
mean population of the \texttt{Arriving Requests} place reaches
approximately $347.67$ requests. The simultaneous increase in this
population and in the mean system response time indicates that request
accumulation at the input of the model becomes a major source of delay
near saturation. The same behavior is reflected in the system
population, shown in Fig.~\ref{fig:spr_capacity_results}
(\subref{fig:spr_capacity_population}). The number of requests present
in the system remains comparatively small throughout the stable
operating region but grows rapidly as the workload approaches the
capacity boundary. At $\lambda=15$ req/s, the capacity-limited
model contains an average of approximately $416.28$ requests. In
contrast, the unlimited SPR baseline predicts only about $5$ requests
at the same arrival rate. This difference shows that the major
degradation near the capacity boundary is associated with request
accumulation and the resulting queueing delay, even while system
throughput remains close to the incoming rate.

System power provides a complementary view of this performance
transition. As shown in Fig.~\ref{fig:spr_capacity_results}
(\subref{fig:spr_capacity_power}), $P_{sys}$ initially increases as
the workload grows because throughput increases while response time
remains relatively low. The metric reaches a maximum value of
approximately $21.1$ at $\lambda=12.5$ req/s. Beyond this point,
the increase in response time outweighs the corresponding throughput
gain, causing system power to decline. Hence, the power curve
indicates that the throughput--latency trade-off begins to deteriorate
before the most severe queue accumulation is observed near the
capacity limit.

Overall, Fig.~\ref{fig:spr_capacity_results} shows that the
capacity-limited SPR model remains close to the ideal baseline at low
and moderate workloads, but its behavior changes substantially in the
high-load region. Throughput remains comparatively stable, whereas
response time, the population of the \texttt{Arriving Requests} place,
and total system population increase rapidly as the available FI
capacity is approached. These results reveal a clear QoS degradation
region that is not captured by the ideal $M/M/\infty$ baseline.
Although SPR conceptually provides processing capacity on a
per-request basis, practical serverless environments operate under
finite resource availability and provider-specific scaling
constraints. The 15-FI limit used in the QPN model is not intended
to reproduce the configuration of a specific commercial platform;
rather, it provides a controlled finite-capacity boundary for
examining the departure from the ideal $M/M/\infty$ behavior. Thus,
the unlimited model serves as an ideal analytical reference, whereas
the capacity-limited QPN model captures the performance degradation
that may arise when available processing capacity becomes constrained.
\subsection{CLS Performance Evaluation}
\label{subsec:cl_results}

The CLS strategy is first evaluated in terms of the dynamic
state-dependent activation of FIs and the corresponding response-time
behavior. The analysis then investigates the effect of distributing a
fixed total nominal concurrency capacity across different server
configurations. The simulations are performed with up to 300,000
samples and 20 repetitions for each evaluated configuration, and the
reported steady-state performance measures are obtained by averaging
the corresponding simulation runs.

\subsubsection{Dynamic QPN Performance Evaluation}
\label{subsubsec:cl_qpn_evaluation}

To examine the dynamic behavior of the CLS strategy, the experiment considers three FIs
($FI_1$--$FI_3$),
each with a nominal concurrency capacity of 15 requests. The
activation of
$FI_2$ and $FI_3$
follows the state-dependent mechanism described in
Section~\ref{subsec:model_architectures},
whereby additional processing capacity is made available as the
	currently active capacity becomes insufficient.
For this dynamic experiment, the reference $H_2$ parameterization
defined in Section~\ref{subsec:common_setup} is used, corresponding
to a mean arrival rate of $3.75$ req/s. A 200-second simulation is
conducted under this workload.

\begin{figure*}[pos=H]
	\centering
	
	\begin{subfigure}[t]{0.48\textwidth}
		\centering
		\includegraphics[
		width=\linewidth,
		keepaspectratio
		]{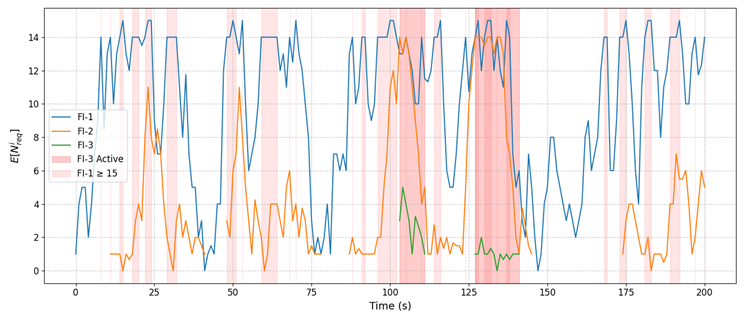}
		\caption{Dynamic evolution of concurrent executions under CLS scaling.}
		\label{fig:cl_exec_time}
	\end{subfigure}
	\hfill
	\begin{subfigure}[t]{0.48\textwidth}
		\centering
		\includegraphics[
		width=\linewidth,
		keepaspectratio
		]{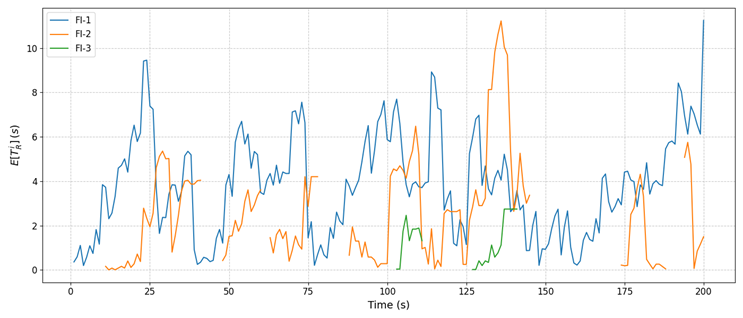}
		\caption{Response-time evolution of FIs under CLS scaling.}
		\label{fig:cl_rt_time}
	\end{subfigure}
	
	\caption{Dynamic performance of the CLS model under the
		$H_2$ workload: (a) concurrent executions over time
		and (b) response-time evolution of the FIs.}
	\label{fig:cl_dynamic_results}
\end{figure*}

The resulting dynamic behavior is summarized in
Fig.~\ref{fig:cl_dynamic_results}. As shown in
Fig.~\ref{fig:cl_dynamic_results}
(\subref{fig:cl_exec_time}), $FI_1$ initially handles most of the
incoming workload and repeatedly operates close to its concurrency
limit. As its available capacity decreases, subsequent requests are
admitted to $FI_2$. During periods of higher demand, particularly
around $t=100$--$145$ seconds, the concurrent load on $FI_2$ increases
substantially and $FI_3$ is also activated. This behavior illustrates
the state-dependent expansion of concurrency capacity when the
currently active FIs are no longer sufficient to absorb the
incoming workload.

The corresponding response-time behavior is presented in
Fig.~\ref{fig:cl_dynamic_results}
(\subref{fig:cl_rt_time}). The response time of $FI_1$ varies as its
concurrency level changes and exhibits several peaks during periods of
increased load. When $FI_2$ begins to absorb additional requests, its
response time initially remains relatively low but increases as its
own concurrency capacity becomes more heavily utilized. The most
pronounced increase occurs during the high-load interval around
$t=125$--$145$ seconds, where the response time of $FI_2$ reaches
approximately $11$ seconds. During the same interval, $FI_3$ becomes
active and provides additional capacity, although its response time
also increases as additional requests are admitted to it.

Together, these results show that the QPN model captures the
time-dependent use of concurrency capacity as the workload varies.
When the capacity of an active FI is exhausted, subsequent requests
can be admitted to the next available FI, thereby increasing the
available processing capacity. Sustained workload bursts can, however,
progressively occupy the concurrency capacity of multiple FIs and lead
to increased response time.

\subsubsection{Wide vs. Deep Concurrency Configurations}
\label{subsubsec:cl_wide_deep}

To examine the effect of concurrency organization on CLS performance,
four configurations (M1--M4) are evaluated while keeping
the total concurrency-token budget fixed at 12 tokens.

\begin{table}[pos=H]
	\caption{CLS configurations with a fixed total concurrency-token
		budget of 12 tokens.}
	\label{tbl:cl_configs_4_6}
	\centering
	\begin{tabular}{@{} l c c c @{}}
		\toprule
		Model & FIs & Tokens per FI & Total tokens
		\tabularnewline
		\midrule
		M1 & 2 & 6 & 12 \tabularnewline
		M2 & 3 & 4 & 12 \tabularnewline
		M3 & 4 & 3 & 12 \tabularnewline
		M4 & 6 & 2 & 12 \tabularnewline
		\bottomrule
	\end{tabular}
\end{table}

As summarized in
Table~\ref{tbl:cl_configs_4_6}, the configurations range from a
deep structure, with fewer FIs and more concurrency tokens per
FI, to wider structures, with more FIs and fewer tokens per FI.
Thus, the quantity held constant across the four configurations
	is the total number of concurrency tokens, while the number of FIs
	and the distribution of those tokens among the FIs vary. The
	service-time characteristics of each FI remain unchanged across
	all configurations.

\begin{figure*}[pos=H]
	\centering
	
	\begin{subfigure}[t]{0.48\textwidth}
		\centering
		\includegraphics[
		width=\linewidth,
		keepaspectratio
		]{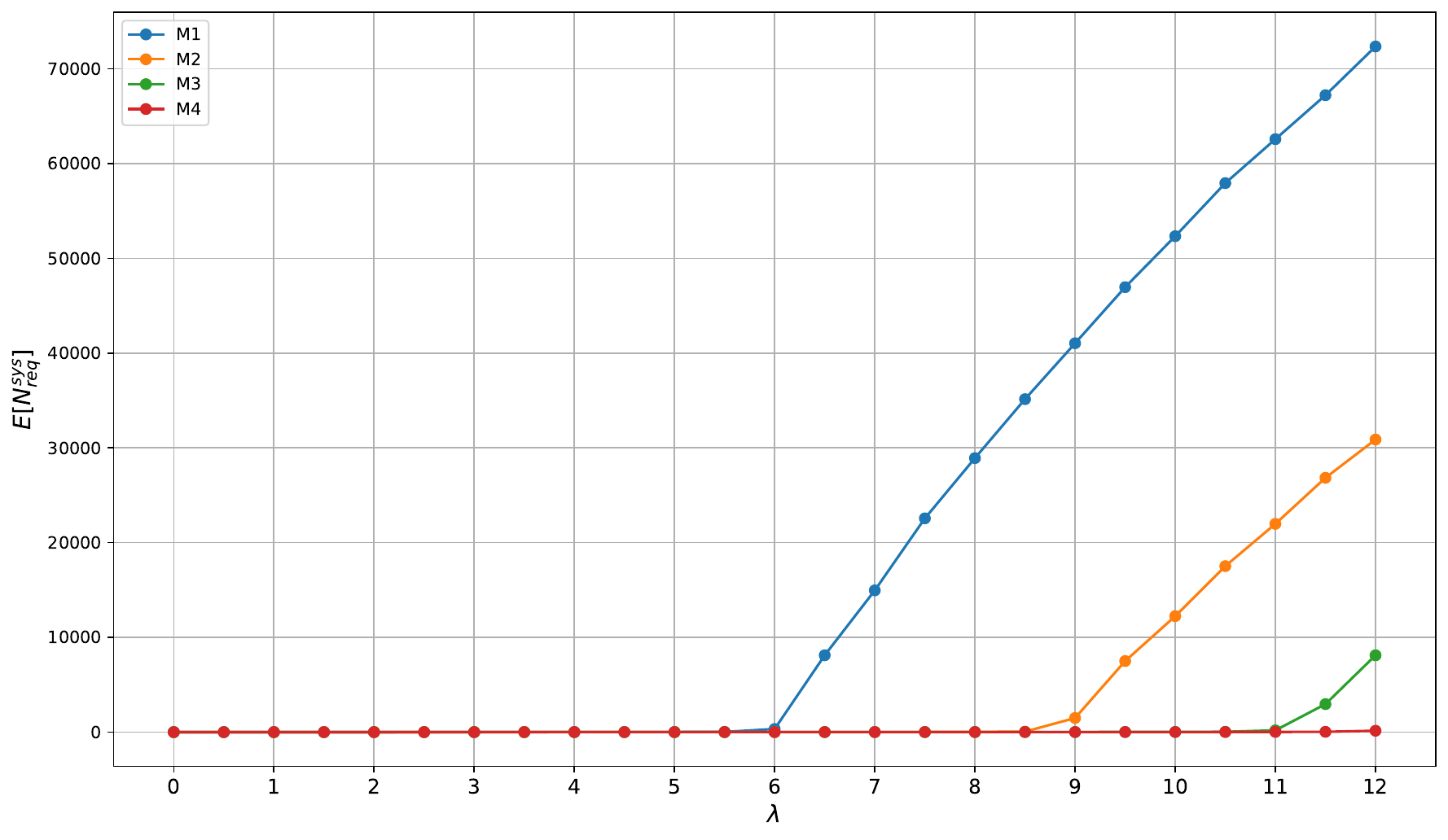}
		\caption{System population.}
		\label{fig:cl_pop_cmp}
	\end{subfigure}
	\hfill
	\begin{subfigure}[t]{0.48\textwidth}
		\centering
		\includegraphics[
		width=\linewidth,
		keepaspectratio
		]{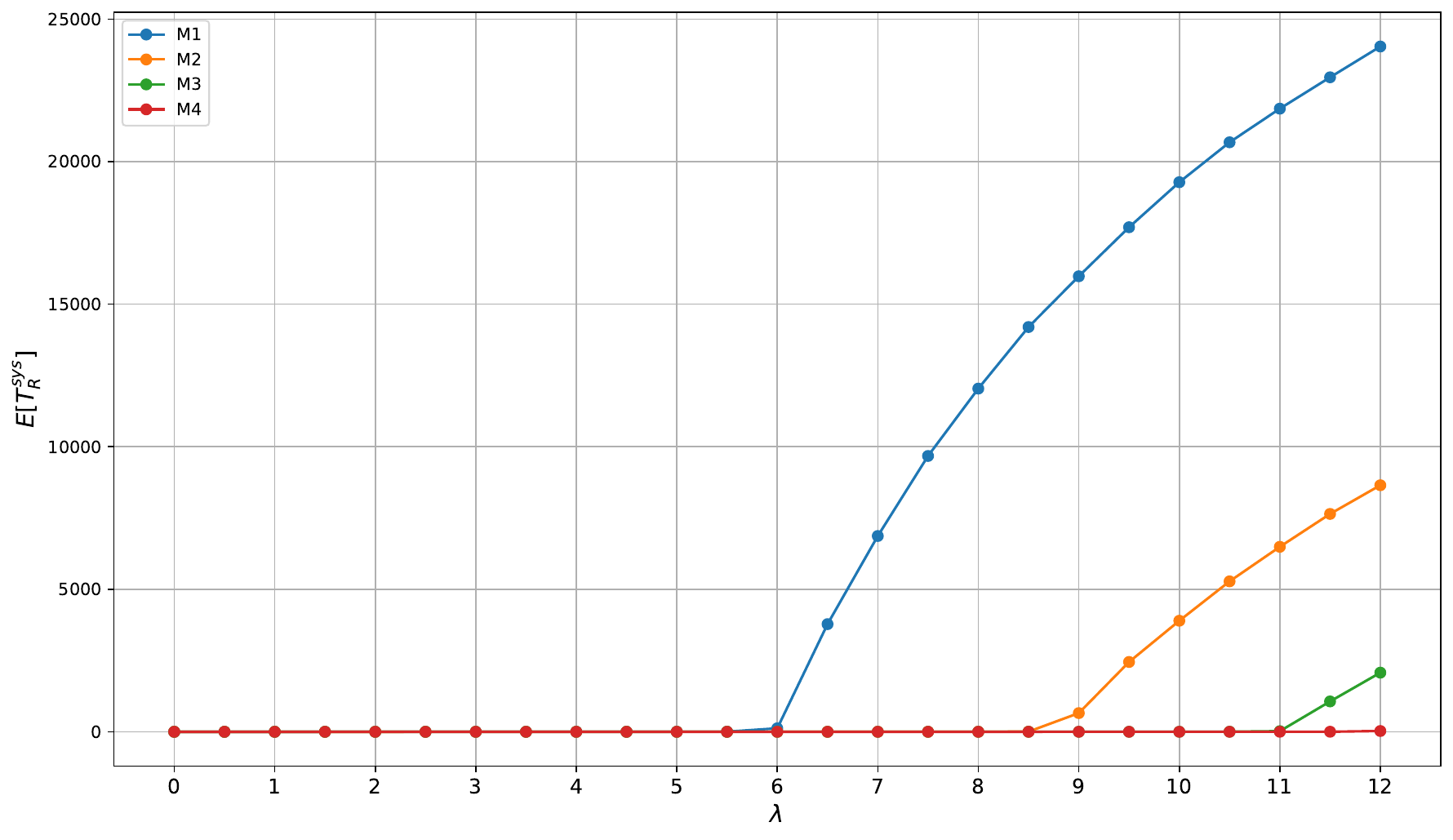}
		\caption{Mean system response time.}
		\label{fig:cl_rt_cmp}
	\end{subfigure}
	
	\vspace{0.25cm}
	
	\begin{subfigure}[t]{0.48\textwidth}
		\centering
		\includegraphics[
		width=\linewidth,
		keepaspectratio
		]{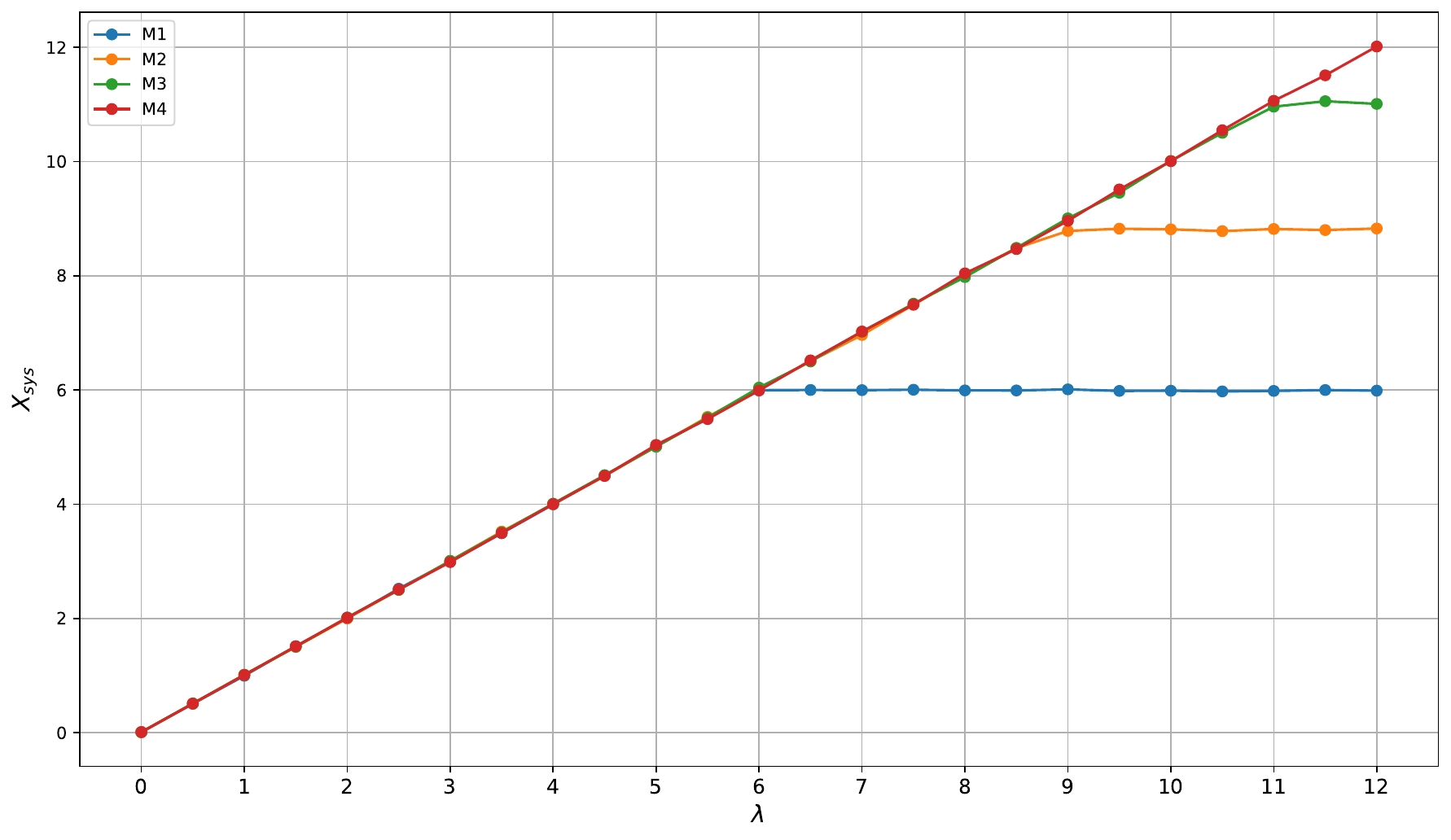}
		\caption{System throughput.}
		\label{fig:cl_thr_cmp}
	\end{subfigure}
	\hfill
	\begin{subfigure}[t]{0.48\textwidth}
		\centering
		\includegraphics[
		width=\linewidth,
		keepaspectratio
		]{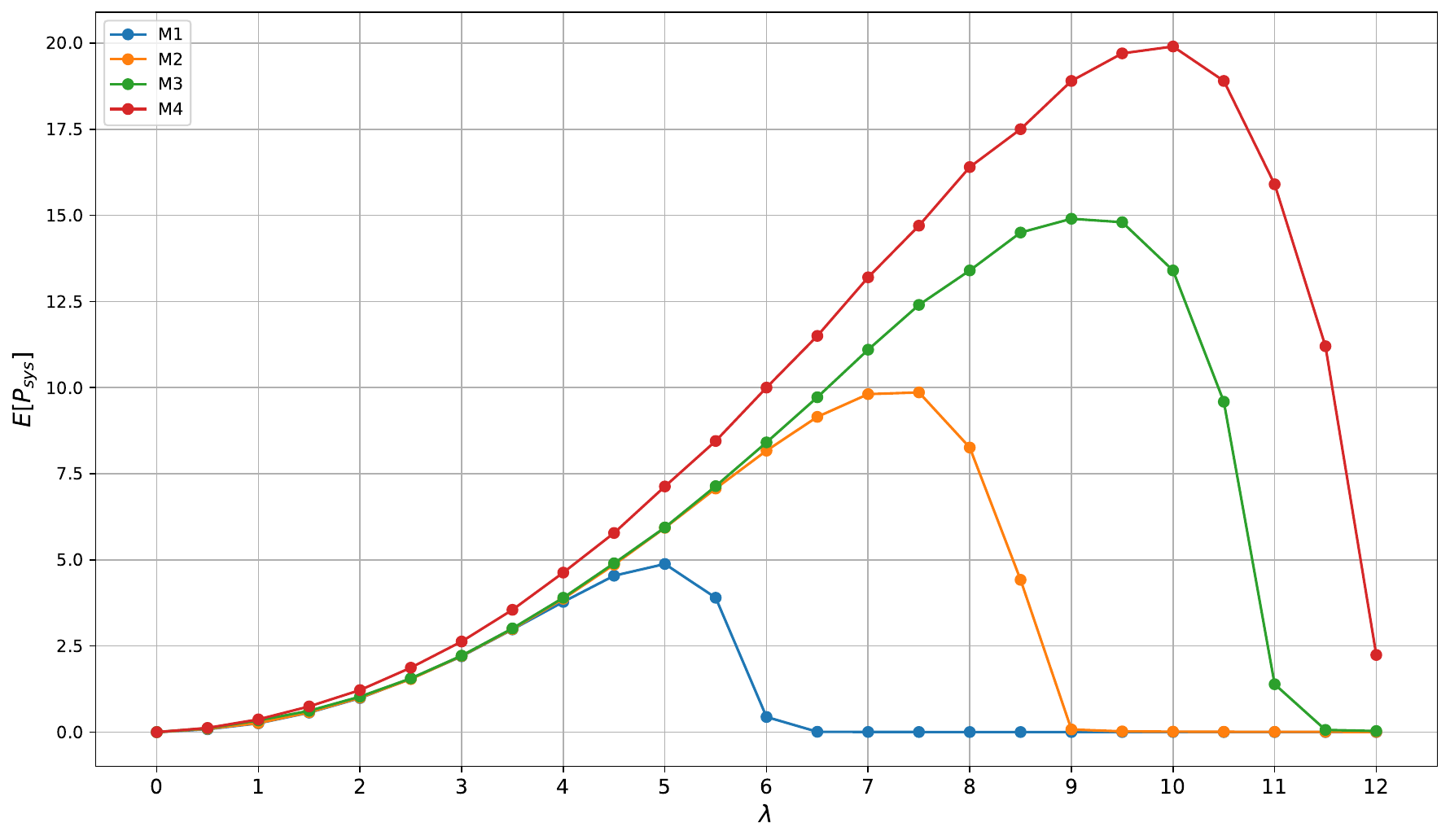}
		\caption{System power.}
		\label{fig:cl_power_cmp}
	\end{subfigure}
	
	\caption{Performance comparison of the CLS configurations with
		a fixed total concurrency-token budget:
		(a) system population,
		(b) mean system response time, (c) system throughput, and
		(d) system power.}
	\label{fig:cl_wide_deep_results}
\end{figure*}

The comparative results are presented in
Fig.~\ref{fig:cl_wide_deep_results}. The system-population and
response-time results in
Fig.~\ref{fig:cl_wide_deep_results}(a) and
Fig.~\ref{fig:cl_wide_deep_results}(b) show that the deep
configurations experience QoS degradation at considerably lower
arrival rates. M1 begins to accumulate requests rapidly from around
$\lambda=6$, whereas the onset of substantial accumulation is
progressively delayed in M2 and M3. M4 exhibits the most controlled
behavior over the evaluated range. At $\lambda=12$ req/s, the system
populations are approximately $72{,}360$, $30{,}880$, $8{,}110$, and
$145$ requests for M1--M4, respectively. The response-time curves
follow the same general pattern, indicating that a wider distribution
of concurrency capacity delays the onset of severe queueing.

The throughput results in
Fig.~\ref{fig:cl_wide_deep_results}(c) provide a consistent view of
this behavior. M1 reaches an effective throughput ceiling of
approximately $6$ req/s, while M2 and M3 sustain progressively
higher arrival rates before their throughput begins to level off.
In contrast, M4 follows the offered workload over almost the entire
evaluated range.
The observed behavior can be explained by the number of FIs 	available for request processing:
The service rate of each FI is 3 req/s.
With 2, 3, 4, and 6 FIs, M1--M4 have aggregate nominal
	processing rates of approximately 6, 9, 12, and 18 req/s,
	respectively.
As the offered load approaches or exceeds the available
	processing capacity, requests accumulate, response time increases,
	and throughput eventually saturates.
The higher capacities of wider configurations postpone
	these effects to higher arrival rates.
Because the workload is bursty, temporary request accumulation
may occur before these nominal processing limits are reached.

System power, shown in
Fig.~\ref{fig:cl_wide_deep_results}(d),
exhibits a corresponding shift in its peak as the number
	of FIs increases.
M1 reaches a peak value of approximately $5$ around $\lambda=5$,
followed by M2 with a peak of about $9.86$ at $\lambda=7.5$
and M3 with approximately $14.9$ at $\lambda=9$.
M4 achieves the highest peak, approximately $19.9$ at
$\lambda=10$, and maintains favorable
throughput--response-time behavior over a wider workload
range before the metric declines.
Overall, deeper configurations concentrate the same token
budget on fewer processing stations, while wider
configurations distribute it across more parallel FIs.
M4 therefore exhibits the most favorable performance
among the evaluated CLS configurations.

\begin{figure*}[pos=H]
	\centering
	
	\begin{subfigure}[t]{\textwidth}
		\centering
		\includegraphics[
		width=\linewidth,
		keepaspectratio
		]{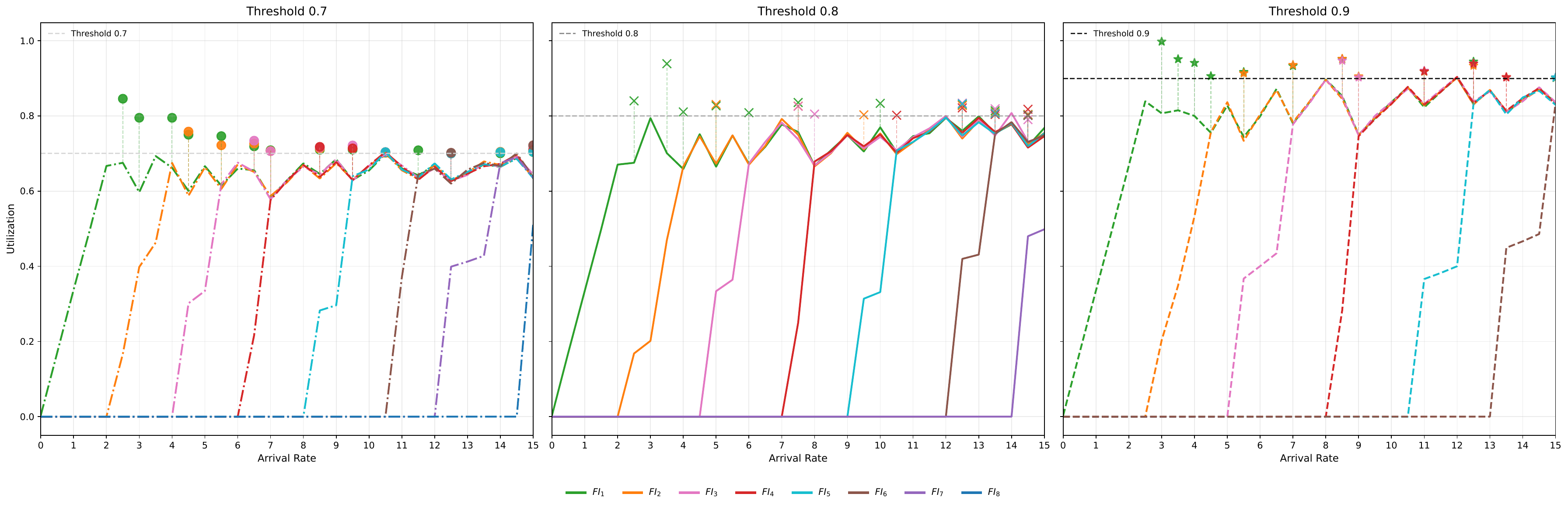}
		\caption{Utilization behavior for
			$U_{\mathrm{th}}=0.7$, $0.8$, and $0.9$.}
		\label{fig:aspm_thr_cmp}
	\end{subfigure}
	
	\vspace{0.35cm}
	
	\begin{subfigure}[t]{\textwidth}
		\centering
		\includegraphics[
		width=\linewidth,
		keepaspectratio
		]{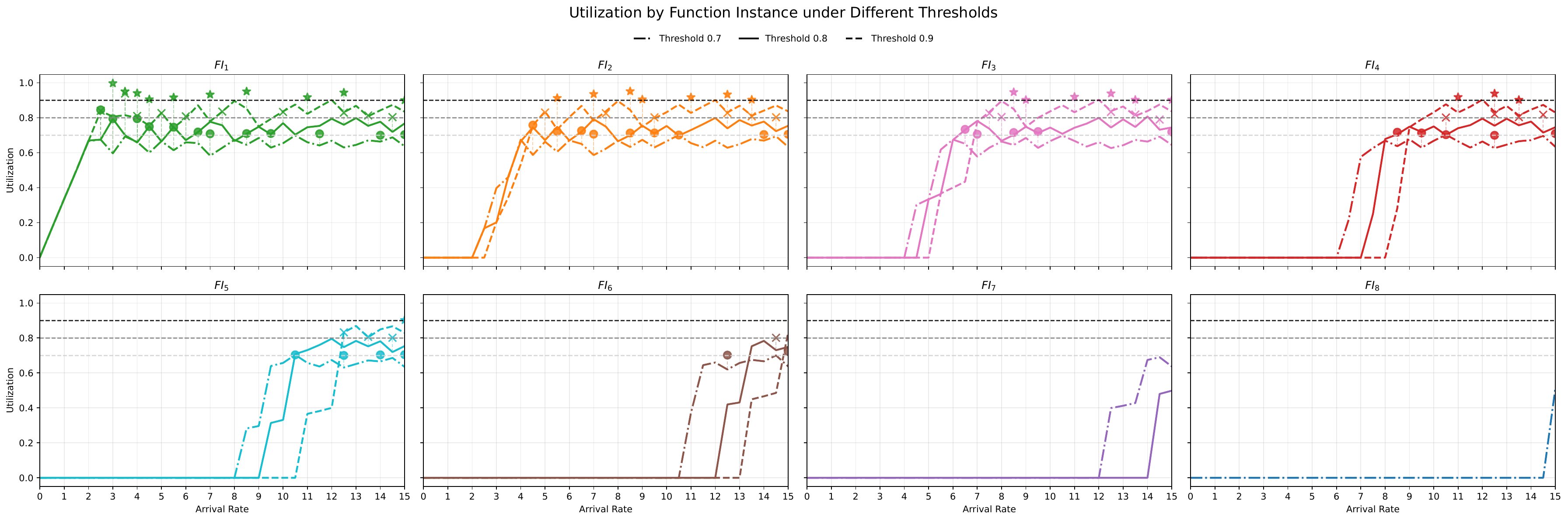}
		\caption{Per-FI utilization and activation behavior across
			the evaluated thresholds. Circles, crosses, and stars denote
			$U_{\mathrm{th}}=0.7$, $0.8$, and $0.9$, respectively.}
		\label{fig:aspm_thr_perfn}
	\end{subfigure}
	
	\caption{Sensitivity of MBS scaling behavior to the utilization
		threshold: (a) utilization profiles across the three evaluated
		thresholds and (b) activation behavior of individual FIs.}
	\label{fig:aspm_threshold_results}
\end{figure*}

\subsection{MBS Performance Evaluation}
\label{subsec:aspm_results}

Section~\ref{subsubsec:aspm_threshold} first examines the sensitivity
of MBS to different utilization thresholds and their effect on scaling
behavior. Section~\ref{subsubsec:aspm_routing} then investigates the
influence of different routing policies on request distribution and
system performance. The resulting configuration is subsequently used
in Section~\ref{subsec:comparative_results} for comparison with the SPR
and CLS strategies.

The JMT-based MBS simulations are performed with up to
1,000,000 samples and five repetitions for each evaluated setting.
The reported performance measures are obtained by averaging the
corresponding simulation runs.

\subsubsection{Utilization Threshold Sensitivity}
\label{subsubsec:aspm_threshold}

The utilization threshold determines when MBS makes additional FIs
available as the workload increases. To examine its effect on scaling
behavior, three threshold values,
$U_{\mathrm{th}}\in\{0.7,0.8,0.9\}$, are evaluated under the same
workload conditions. Fig.~\ref{fig:aspm_threshold_results}(a)
compares the utilization profiles obtained for the three settings.
A lower threshold causes new FIs to become active earlier, whereas
increasing the threshold allows the currently active FIs to operate
at higher utilization levels before additional capacity is introduced.

This shift in activation behavior is more clearly visible in
Fig.~\ref{fig:aspm_threshold_results}(b). With
$U_{\mathrm{th}}=0.7$, capacity is expanded earlier as the arrival
rate increases, leaving greater utilization headroom for subsequent
workload bursts. At $U_{\mathrm{th}}=0.8$, FI activation is delayed,
while the $0.9$ setting keeps the active FIs closer to their
utilization limit before further capacity is introduced. The per-FI
results also show that increasing the threshold progressively shifts
the activation of later FIs toward higher arrival rates. These results
illustrate the trade-off introduced by the utilization threshold.
Lower values provide a larger operating margin against sudden
workload increases but require earlier provisioning of additional
FIs. Higher values make greater use of the capacity already available,
while leaving less headroom when the workload changes abruptly.

\begin{figure*}[pos=H]
	\centering
	
	\begin{subfigure}[t]{0.48\textwidth}
		\centering
		\includegraphics[
		width=\linewidth,
		keepaspectratio
		]{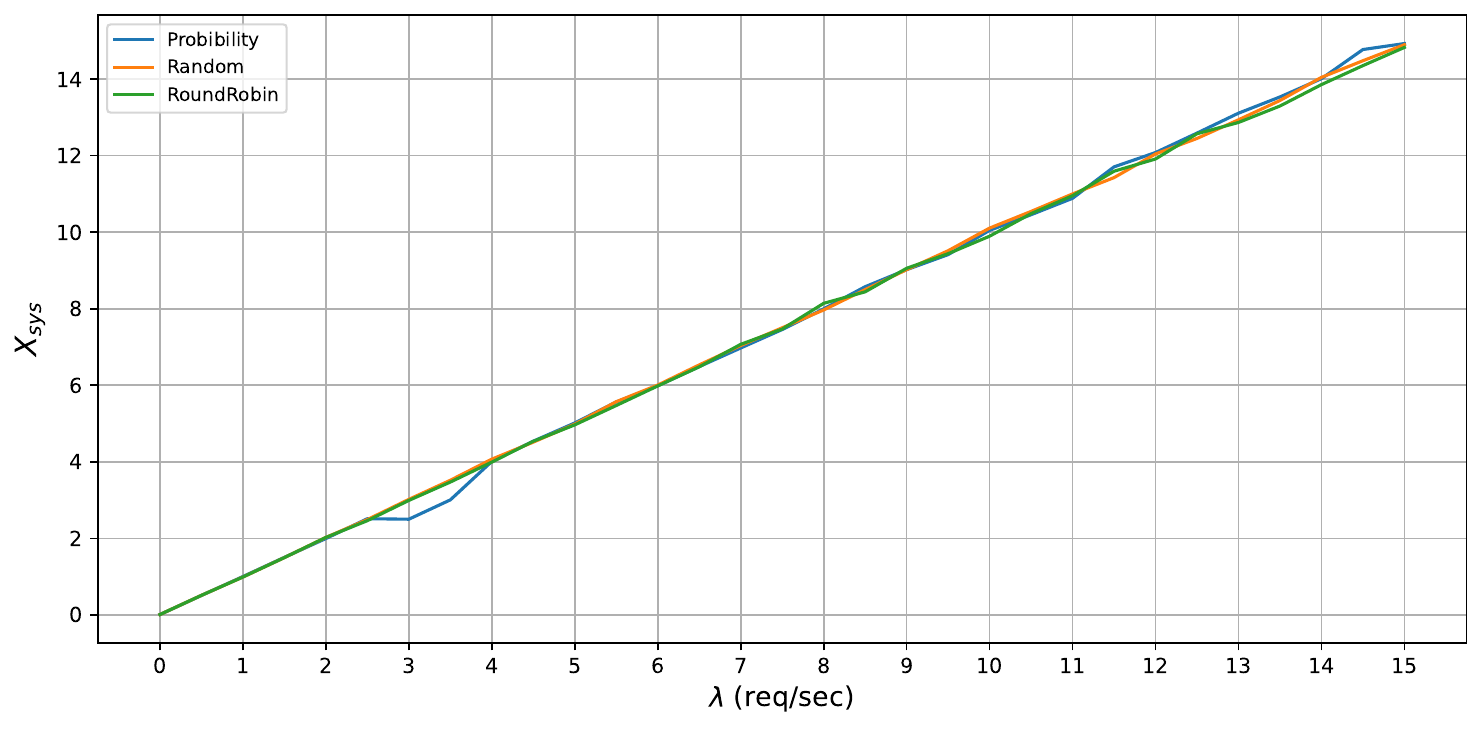}
		\caption{System throughput.}
		\label{fig:aspm_route_thr}
	\end{subfigure}
	\hfill
	\begin{subfigure}[t]{0.48\textwidth}
		\centering
		\includegraphics[
		width=\linewidth,
		keepaspectratio
		]{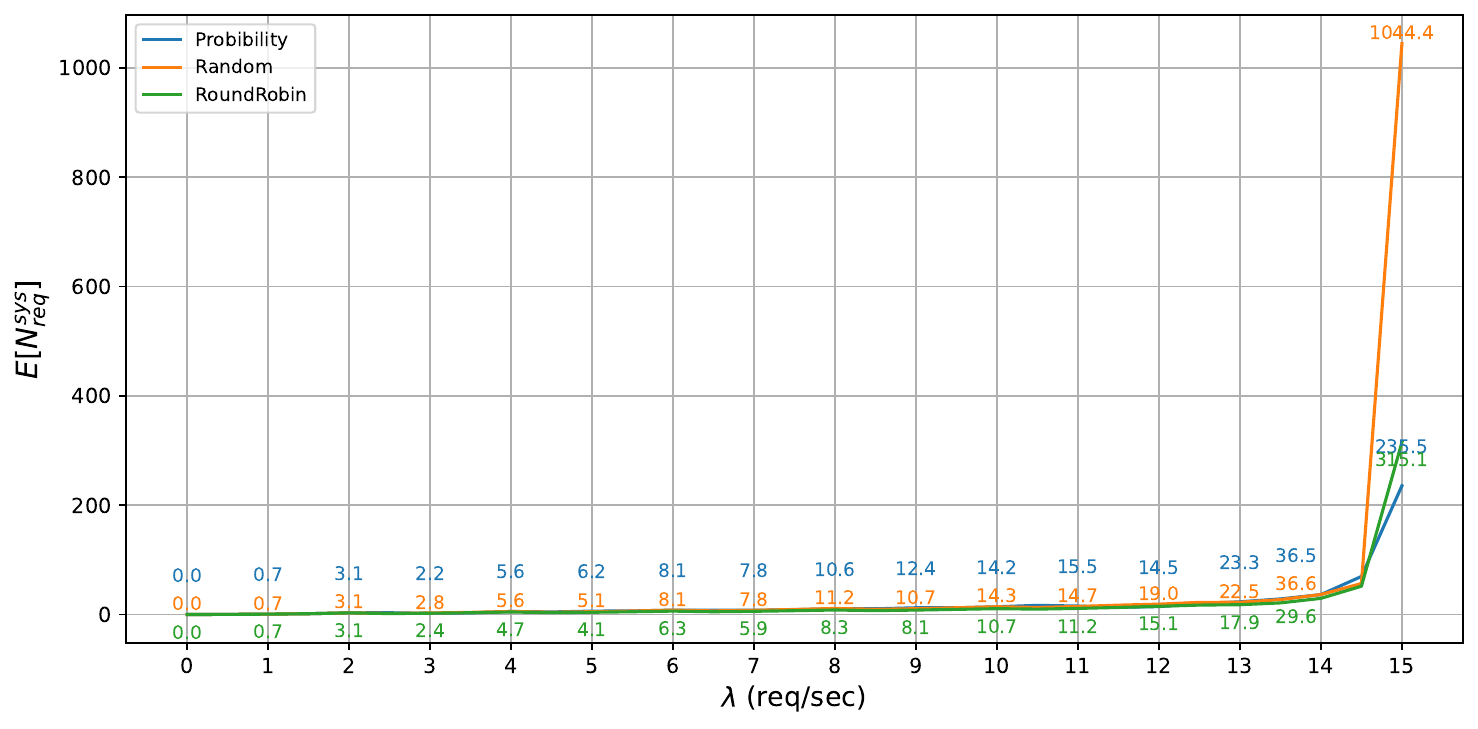}
		\caption{System population.}
		\label{fig:aspm_route_pop}
	\end{subfigure}
	
	\vspace{0.25cm}
	
	\begin{subfigure}[t]{0.48\textwidth}
		\centering
		\includegraphics[
		width=\linewidth,
		keepaspectratio
		]{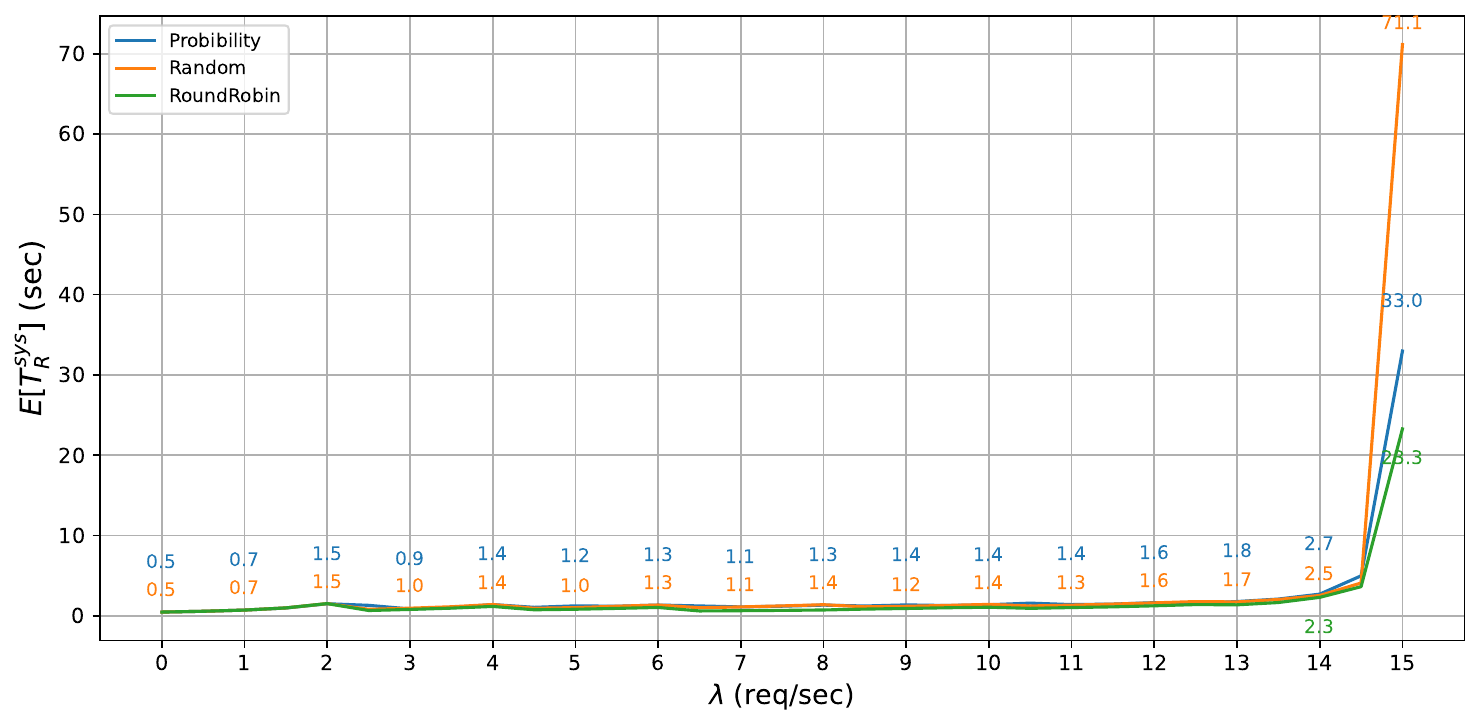}
		\caption{Mean system response time.}
		\label{fig:aspm_route_rt}
	\end{subfigure}
	\hfill
	\begin{subfigure}[t]{0.48\textwidth}
		\centering
		\includegraphics[
		width=\linewidth,
		keepaspectratio
		]{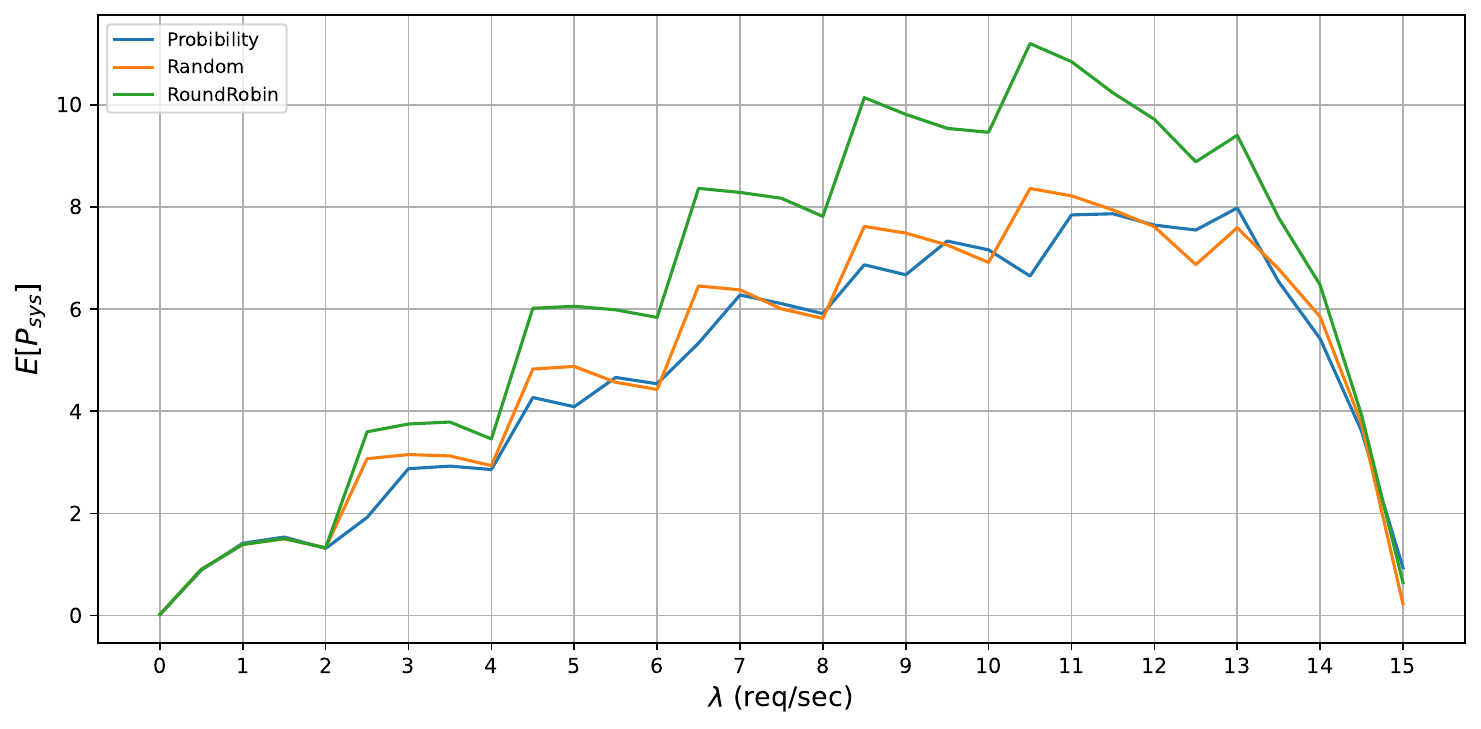}
		\caption{System power.}
		\label{fig:aspm_route_power}
	\end{subfigure}
	
	\caption{Sensitivity of MBS performance to the routing policy
		under $U_{\mathrm{th}}=0.7$: (a) system throughput,
		(b) system population, (c) mean system response time, and
		(d) system power.}
	\label{fig:aspm_routing_results}
\end{figure*}

\subsubsection{Routing Policy Sensitivity}
\label{subsubsec:aspm_routing}

Following the threshold analysis, the utilization threshold is fixed
at $U_{\mathrm{th}}=0.7$ to examine the effect of request routing
under the same scaling threshold.
Three routing policies are considered: \textit{probabilistic},
\textit{random}, and \textit{round-robin}. Their performance is
evaluated in terms of system throughput, population, response time,
and system power, as shown in
Fig.~\ref{fig:aspm_routing_results}.

The throughput curves in
Fig.~\ref{fig:aspm_routing_results}(a) remain close across the
three policies over the evaluated workload range. This indicates
that changing the routing policy has only a limited effect on the
overall request completion rate. Differences become more apparent
in the system population shown in
Fig.~\ref{fig:aspm_routing_results}(b). As the arrival rate
increases, round-robin maintains a lower request population than the
random and probabilistic policies, indicating a more balanced
distribution of requests among the active FIs.

A similar pattern appears in the response-time results in
Fig.~\ref{fig:aspm_routing_results}(c). The three policies perform
similarly at low arrival rates, but their behavior diverges as the
workload approaches the upper end of the evaluated range. At
$\lambda=15$ req/s, the mean response time reaches approximately
$23.37$ seconds for round-robin, compared with $33.0$ seconds for
probabilistic routing and $71.1$ seconds for random routing. Thus,
round-robin limits the growth in delay more effectively under the
considered high-load conditions.

Fig.~\ref{fig:aspm_routing_results}(d) shows the corresponding
system-power behavior. Since system power combines throughput and
response time, the differences among the routing policies become more
visible as response time begins to increase. The curves reach their
highest region at approximately
$\lambda=10.5$ req/s
and decline thereafter as the increase in response time becomes more
pronounced.

The local variations in system population and response time are
	mainly associated with threshold-based FI activation and request
	redistribution. Activating another FI temporarily reduces queueing by
	distributing incoming requests across more active FIs, after which
	these measures increase again as the workload grows. The bursty
$H_2$
workload can further contribute to these variations, which are
	also reflected in system power through changes in response time.
Among the evaluated routing policies, round-robin provides the
	most favorable behavior in the higher-load
	region. Based on these results, round-robin routing with
$U_{\mathrm{th}}=0.7$
is adopted for the MBS configuration used in the comparative
	evaluation of
Section~\ref{subsec:comparative_results}.

\begin{figure*}[pos=H]
	\centering
	
	\begin{subfigure}[t]{0.48\textwidth}
		\centering
		\includegraphics[
		width=\linewidth,
		keepaspectratio
		]{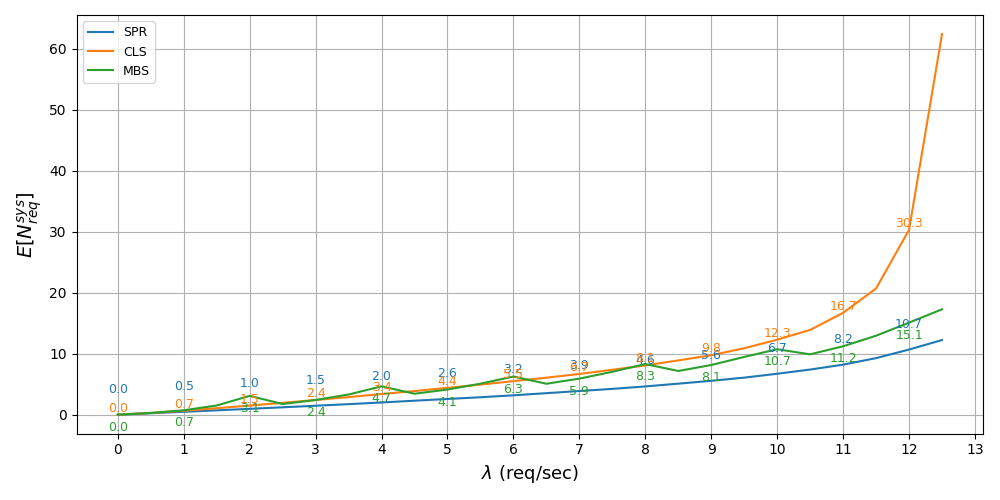}
		\caption{System population.}
		\label{fig:cmp_pop}
	\end{subfigure}
	\hfill
	\begin{subfigure}[t]{0.48\textwidth}
		\centering
		\includegraphics[
		width=\linewidth,
		keepaspectratio
		]{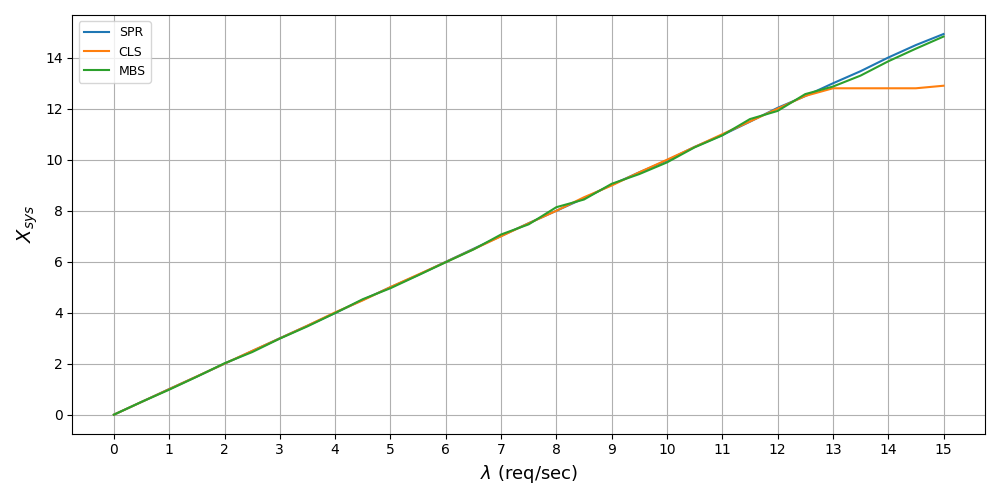}
		\caption{System throughput.}
		\label{fig:cmp_thr}
	\end{subfigure}
	
	\vspace{0.25cm}
	
	\begin{subfigure}[t]{0.48\textwidth}
		\centering
		\includegraphics[
		width=\linewidth,
		keepaspectratio
		]{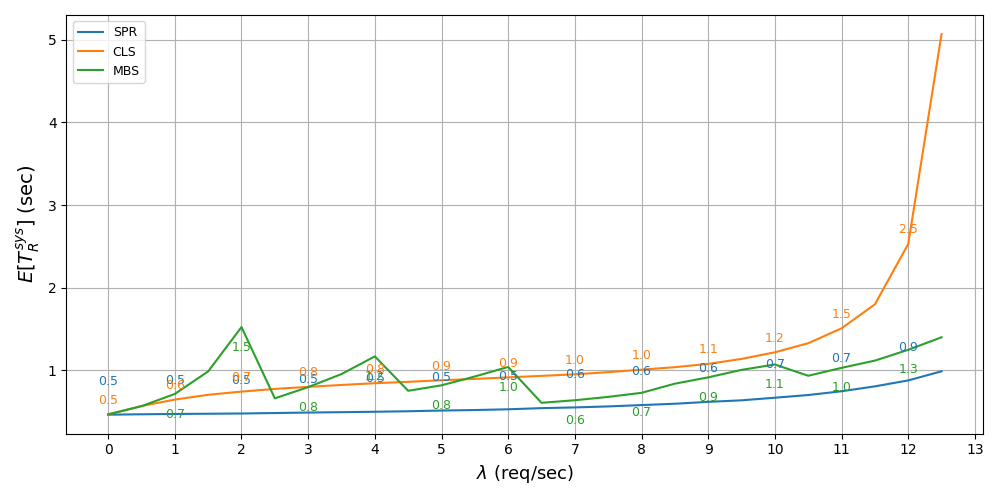}
		\caption{Mean system response time.}
		\label{fig:cmp_rt}
	\end{subfigure}
	\hfill
	\begin{subfigure}[t]{0.48\textwidth}
		\centering
		\includegraphics[
		width=\linewidth,
		keepaspectratio
		]{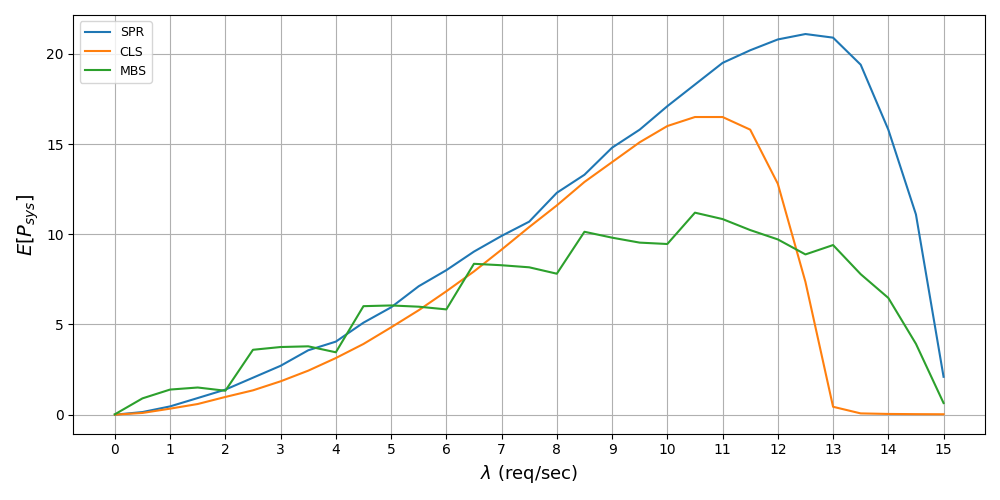}
		\caption{System power.}
		\label{fig:cmp_power}
	\end{subfigure}
	
	\caption{Comparative performance of SPR, CLS, and MBS:
		(a) system population, (b) system throughput,
		(c) mean system response time, and (d) system power.}
	\label{fig:cmp_all}
\end{figure*}

\subsection{Comparative Evaluation of Autoscaling Strategies}
\label{subsec:comparative_results}

The SPR, CLS, and MBS strategies are compared under the 
workload conditions defined in Section~\ref{subsec:common_setup} to
examine how their different scaling mechanisms affect system behavior
as the offered load increases. The mean arrival rate is varied from
$\lambda=0$ to $15$ req/s over the evaluated workload range. For this
investigation, SPR can activate up to 15 FIs, CLS is configured with
five FIs and three concurrency tokens per FI, and MBS uses the
utilization threshold $U_{\mathrm{th}}=0.7$ with round-robin routing.
The comparison considers system population, throughput, response time,
and system power, as summarized in Fig.~\ref{fig:cmp_all}.
Although all simulations cover the complete workload range up to
$\lambda=15$ req/s,
the system-population and mean-response-time plots are displayed
	only up to
$\lambda=12.5$ req/s. At higher arrival rates, the corresponding values,
	particularly for CLS, increase sharply; displaying these points on the
	same scale would compress the lower-load region and obscure the
	differences among the strategies. The high-load results are therefore
	reported explicitly in the following discussion.

The system-population results in
Fig.~\ref{fig:cmp_all}(\subref{fig:cmp_pop}) show similar behavior
among the three strategies at low workloads, but their responses
increasingly diverge as the arrival rate grows. SPR maintains the
lowest population over most of the
low-to-moderate workload range
($0 \leq \lambda \leq 12.5$ req/s),
reflecting its ability to assign processing capacity directly to
incoming requests while sufficient FIs remain available.
CLS begins to accumulate requests earlier as its fixed
	concurrency configuration increasingly restricts request admission
	across the available FIs.
This effect becomes particularly pronounced near the upper end of
the workload range. MBS exhibits somewhat higher population than SPR
under moderate loads because capacity expansion is governed by the
utilization threshold rather than by each individual arrival;
however, when the utilization threshold is reached, additional
	processing capacity is activated, which limits queue growth more
	effectively as the system approaches its capacity boundary.
Based on the complete simulation results, at
$\lambda=15$ req/s,
the system population reaches approximately
416.3 requests for SPR and 15,300 requests for CLS, compared with
approximately 315.1 requests for MBS.
The substantially larger CLS population at this workload reflects
	the severe request accumulation occurring beyond the range displayed
	in
Fig.~\ref{fig:cmp_all}(\subref{fig:cmp_pop}).

A different pattern is observed for system throughput in
Fig.~\ref{fig:cmp_all}(\subref{fig:cmp_thr}). All three strategies
closely follow the offered workload through the low- and moderate-load
regions, and their differences remain small until the arrival rate
approaches the available processing capacity. SPR sustains a
throughput of approximately $14.92$ req/s at $\lambda=15$ req/s,
while MBS reaches approximately $14.82$ req/s. CLS begins to level
off earlier and provides approximately $12.9$ req/s at the same
arrival rate.
Thus, SPR and MBS preserve most of the incoming throughput near
	the upper workload limit, whereas the fixed CLS configuration reaches
	its effective processing limit earlier as the workload increases.

The response-time results provide a clearer indication of the
different QoS behavior of the three strategies. As illustrated in
Fig.~\ref{fig:cmp_all}(\subref{fig:cmp_rt}), SPR provides the lowest
response time through most of the low-to-moderate workload range,
while MBS remains relatively close to it. CLS shows a more noticeable
increase as its concurrency capacity becomes progressively occupied.
The differences become substantially larger near the maximum evaluated
load.
Based on the complete simulation results beyond the displayed
	range of

Fig.~\ref{fig:cmp_all}(\subref{fig:cmp_rt}),
at
$\lambda=15$ req/s,
the mean response time rises to approximately $62.4$ seconds for SPR
and $3240$ seconds for CLS, whereas MBS limits the corresponding value
to approximately $23.27$ seconds. These results indicate that SPR
provides favorable latency while sufficient per-request capacity is
available, but its response time deteriorates rapidly when the finite
FI boundary is approached. The utilization-driven activation used by
MBS introduces additional capacity before this condition becomes as
severe, resulting in a more gradual increase in delay.

Finally,
Fig.~\ref{fig:cmp_all}(\subref{fig:cmp_power}) compares system power,
defined as the ratio
$P_{\mathrm{sys}}=X_{\mathrm{sys}}/E[T_R^{\mathrm{sys}}]$.
SPR achieves the highest peak value, approximately $21.1$, around
$\lambda=12.5$ req/s, reflecting its combination of high
throughput and low response time in this operating region. CLS
reaches a lower peak of approximately $16.5$ around
$\lambda=10.5$--$11$ req/s and then declines sharply as its
response time increases. MBS has a lower peak, approximately $11.19$,
but exhibits a more gradual change as the workload increases.
Consequently, the power results should not be interpreted as showing
a single strategy that dominates across all operating conditions.
Rather, they reinforce the trade-off observed in the other metrics:
SPR provides the strongest performance under low-to-moderate loads,
CLS is more sensitive to concurrency-capacity exhaustion, and MBS
sacrifices some peak performance in favor of more controlled behavior
near the capacity limit.




\section{Conclusion and Future Work}
\label{sec:conclusion}

This study examined the behavior of the three autoscaling strategies under dynamic
and bursty workloads. Analytical queueing models and QPN models were
used to capture their different scaling mechanisms, while an $H_2$
arrival process was used to represent workload variability.

For SPR, the $M/M/\infty$ model provided an analytical reference for
the corresponding simulation model. Under finite processing capacity,
SPR maintained low response time over much of the evaluated workload
range, but its performance deteriorated rapidly as the available
FI capacity was approached.

The CLS results showed that performance depends not only on the total
concurrency-token budget, but also on how this budget is distributed
across FIs. In the wide-versus-deep experiments, distributing the same
token budget across a larger number of independent FIs delayed queue
buildup and provided more favorable behavior under increasing load.

For MBS, the utilization threshold determined the timing of FI
activation, while the routing policy affected how requests were
distributed among active FIs. Among the evaluated routing policies,
round-robin produced the most controlled behavior. The configuration
with $U_{\mathrm{th}}=0.7$ was used in the final comparison with SPR
and CLS.

The comparative results show that the three strategies exhibit
different trade-offs. SPR provides strong responsiveness while
sufficient FI capacity is available, whereas CLS becomes more
sensitive to concurrency exhaustion as the workload increases. MBS
provides a more gradual response near the capacity boundary through
utilization-driven FI activation. The results also show that the
variability introduced by the $H_2$ workload can lead to queue growth
and QoS degradation before nominal capacity is fully reached.
Accordingly, the choice of autoscaling strategy depends on the desired
balance between responsiveness, concurrency control, available
capacity, and robustness to workload variability.

Future work can extend the proposed framework to heterogeneous
workloads and FIs with different service demands and resource
characteristics. Shared external resources, such as databases,
storage services, and external APIs, can also be incorporated into
the QPN models to capture contention beyond individual FIs.
In addition, resource cost can be incorporated     to support
multi-objective scaling decisions that balance QoS and resource usage,
while predictive mechanisms can be investigated to anticipate workload
changes before utilization thresholds are reached.
The evaluated strategies can also be implemented on a
serverless platform such as Knative or OpenFaaS. Comparing platform
measurements with the analytical and simulation results would provide
an empirical assessment of the proposed models and help identify
effects that are not captured by the current abstraction.

\section*{Declaration of Generative AI and AI-assisted Technologies in the Writing Process}

During the preparation of this work the authors used ChatGPT   in order to improve language and readability. After using this tool, the authors reviewed and edited the content as needed and take full responsibility for the content of the manuscript.

\bibliographystyle{elsarticle-num}
\bibliography{cas-refs}

\begin{thebibliography}{10}
\expandafter\ifx\csname url\endcsname\relax
  \def\url#1{\texttt{#1}}\fi
\expandafter\ifx\csname urlprefix\endcsname\relax\def\urlprefix{URL }\fi
\expandafter\ifx\csname href\endcsname\relax
  \def\href#1#2{#2} \def\path#1{#1}\fi

\bibitem{hong2024analysis}
S.~Hong, Y.~Kim, J.~Nam, S.~Kim, On the analysis of inter-relationship between
  auto-scaling policy and {QoS} of {FaaS} workloads, Sensors 24~(12) (2024)
  3774.

\bibitem{tari2024auto}
M.~Tari, M.~Ghobaei-Arani, J.~Pouramini, M.~Ghorbian, Auto-scaling mechanisms
  in serverless computing: A comprehensive review, Computer Science Review 53
  (2024) 100650.

\bibitem{agarwal2024dynamic}
S.~Agarwal, et~al., Dynamic function configuration and its management in
  serverless computing, ACM Computing Surveys (2024).

\bibitem{fan2024puffer}
C.~Fan, et~al., Puffer: Reducing cold-start time in serverless computing, IEEE
  Transactions on Parallel and Distributed Systems (2024).

\bibitem{li2024secure}
Y.~Li, et~al., Lightweight and holistic-scalable serverless secure container
  runtime, IEEE Transactions on Computers (2024).

\bibitem{karamzadeh2024lightweight}
Z.~Karamzadeh, D.~Shameli-Sendi, Reducing cold start delay in serverless
  computing using lightweight virtual machines, Journal of Network and Computer
  Applications (2024).

\bibitem{Alharthi2024Auto-Scaling}
S.~Alharthi, A.~Alshamsi, A.~Alseiari, A.~Alwarafy, Auto-scaling techniques in
  cloud computing: Issues and research directions, Sensors (Basel, Switzerland)
  24 (2024).
\newblock \href {https://doi.org/10.3390/s24175551}
  {\path{doi:10.3390/s24175551}}.

\bibitem{Meng2025CATScaler}
F.~Meng, H.~Dai, G.~Cong, B.~Zhu, H.~Zhao, {CATScaler}: A convolution-augmented
  transformer scaling framework for cloud-native applications, IEEE
  Transactions on Services Computing 18 (2025) 2659--2672.
\newblock \href {https://doi.org/10.1109/tsc.2025.3592383}
  {\path{doi:10.1109/tsc.2025.3592383}}.

\bibitem{Jeong2025ARAScaler}
B.~Jeong, Y.-S. Jeong, {ARAScaler}: Adaptive resource autoscaling scheme using
  {ETimeMixer} for efficient cloud-native computing, IEEE Transactions on
  Services Computing 18 (2025) 72--84.
\newblock \href {https://doi.org/10.1109/tsc.2024.3522815}
  {\path{doi:10.1109/tsc.2024.3522815}}.

\bibitem{Ebrahimi2024Cold}
A.~Ebrahimi, M.~Ghobaei-Arani, H.~Saboohi, Cold start latency mitigation
  mechanisms in serverless computing: Taxonomy, review, and future directions,
  J. Syst. Archit. 151 (2024) 103115.
\newblock \href {https://doi.org/10.1016/j.sysarc.2024.103115}
  {\path{doi:10.1016/j.sysarc.2024.103115}}.

\bibitem{Cai2025SLO-Aware}
X.~Cai, K.~Yan, Y.~Gong, C.~Hu, D.~Cheng, {SLO}-aware instance management with
  queuing-based delay execution, IEEE Transactions on Services Computing 18
  (2025) 4167--4180.
\newblock \href {https://doi.org/10.1109/tsc.2025.3632416}
  {\path{doi:10.1109/tsc.2025.3632416}}.

\bibitem{benedetti2025qlearning}
P.~Benedetti, M.~Femminella, G.~Reali, Management of autoscaling serverless
  functions in edge computing via {Q}-learning, Future Generation Computer
  Systems 175 (2026) 108112.
\newblock \href {https://doi.org/10.1016/j.future.2025.108112}
  {\path{doi:10.1016/j.future.2025.108112}}.

\bibitem{Cai2024Incendio:}
X.~Cai, Q.~Sang, C.~Hu, Y.~Gong, K.~Suo, X.~Zhou, D.~Cheng, Incendio:
  Priority-based scheduling for alleviating cold start in serverless computing,
  IEEE Transactions on Computers 73 (2024) 1780--1794.
\newblock \href {https://doi.org/10.1109/tc.2024.3386063}
  {\path{doi:10.1109/tc.2024.3386063}}.

\bibitem{Quattrocchi2024Autoscaling}
G.~Quattrocchi, E.~Incerto, R.~Pinciroli, C.~Trubiani, L.~Baresi, Autoscaling
  solutions for cloud applications under dynamic workloads, IEEE Transactions
  on Services Computing 17 (2024) 804--820.
\newblock \href {https://doi.org/10.1109/tsc.2024.3354062}
  {\path{doi:10.1109/tsc.2024.3354062}}.

\bibitem{Kulkarni2025Characterizing}
V.~Kulkarni, N.~Reddy, T.~Khare, A.~S. Prasad, C.~Babu, Y.~L. Simmhan,
  Characterizing {FaaS} workflows on public clouds: The good, the bad and the
  ugly, IEEE Transactions on Parallel and Distributed Systems 37 (2025)
  1312--1329.
\newblock \href {https://doi.org/10.1109/tpds.2026.3678606}
  {\path{doi:10.1109/tpds.2026.3678606}}.

\bibitem{tran2024concurrent}
M.-N. Tran, Y.~Kim, Concurrent service auto-scaling for {K}native resource
  quota-based serverless system, Future Generation Computer Systems 160 (2024)
  326--339.
\newblock \href {https://doi.org/10.1016/j.future.2024.06.019}
  {\path{doi:10.1016/j.future.2024.06.019}}.

\bibitem{tran2024optimized}
M.-N. Tran, Y.~Kim, Optimized resource usage with hybrid auto-scaling system
  for {K}native serverless edge computing, Future Generation Computer Systems
  152 (2024) 304--316.

\bibitem{Shafi2025CEMA}
N.~Shafi, M.~Abdullah, W.~Iqbal, F.~Bukhari, {CEMA}: Cost effective
  multi-layered autoscaling for microservice based applications, Journal of
  Network and Computer Applications 242 (2025) 104266.
\newblock \href {https://doi.org/10.1016/j.jnca.2025.104266}
  {\path{doi:10.1016/j.jnca.2025.104266}}.

\bibitem{Zafeiropoulos2022RLAutoscaling}
A.~Zafeiropoulos, E.~Fotopoulou, N.~Filinis, S.~Papavassiliou, Reinforcement
  learning-assisted autoscaling mechanisms for serverless computing platforms,
  Simulation Modelling Practice and Theory 116 (2022) 102461.
\newblock \href {https://doi.org/10.1016/j.simpat.2021.102461}
  {\path{doi:10.1016/j.simpat.2021.102461}}.

\bibitem{khaleq2023intelligent}
A.~A. Khaleq, I.~Ra, Intelligent microservices autoscaling module using
  reinforcement learning, Cluster Computing 26 (2023) 2789--2800.
\newblock \href {https://doi.org/10.1007/s10586-023-03999-8}
  {\path{doi:10.1007/s10586-023-03999-8}}.

\bibitem{mampage2025deep}
A.~Mampage, S.~Karunasekera, R.~Buyya, A deep reinforcement learning based
  algorithm for time and cost optimized scaling of serverless applications,
  Future Generation Computer Systems 173 (2025) 107873.
\newblock \href {https://doi.org/10.1016/j.future.2025.107873}
  {\path{doi:10.1016/j.future.2025.107873}}.

\bibitem{li2025energy}
H.~Li, W.~Rao, B.~Hu, Y.~Tian, J.~Shen, Energy-aware elastic scaling algorithm
  for microservices in {K}ubernetes clouds, Journal of Network and Computer
  Applications 242 (2025) 104218.
\newblock \href {https://doi.org/10.1016/j.jnca.2025.104218}
  {\path{doi:10.1016/j.jnca.2025.104218}}.

\bibitem{abdullah2025dynamic}
M.~Abdullah, R.~Munir, W.~Iqbal, S.~A. Khan, S.~U.~R. Baig, A.~Erradi, Cost and
  performance-effective dynamic {VM} type selection in auto-scaling using
  reinforcement learning, Computing 107~(9) (2025) 177.
\newblock \href {https://doi.org/10.1007/s00607-025-01535-7}
  {\path{doi:10.1007/s00607-025-01535-7}}.

\bibitem{ship2024optimizing}
H.~Ship, E.~Shindin, C.~Wang, D.~Arroyo, A.~Tantawi, Optimizing simultaneous
  autoscaling for serverless cloud computing, in: 2024 IEEE 17th International
  Conference on Cloud Computing (CLOUD), IEEE, 2024, pp. 105--114.
\newblock \href {https://doi.org/10.1109/CLOUD62652.2024.00022}
  {\path{doi:10.1109/CLOUD62652.2024.00022}}.

\bibitem{zafarzade2025capacity}
Z.~Zafarzade, E.~Ataie, Capacity planning of a microservices-based image
  classification application using analytic modeling, Computing 107 (2025) 123.
\newblock \href {https://doi.org/10.1007/s00607-025-01473-4}
  {\path{doi:10.1007/s00607-025-01473-4}}.

\bibitem{wu2025cloudnativesim}
J.~Wu, M.~Xu, Y.~He, K.~Ye, C.~Xu, {C}loud{N}ative{S}im: A toolkit for modeling
  and simulation of cloud-native applications, Software: Practice and
  Experience 55 (2025) 1185--1208.
\newblock \href {https://doi.org/10.1002/spe.3417}
  {\path{doi:10.1002/spe.3417}}.

\bibitem{fe2022performance}
I.~F{\'e}, R.~Matos, J.~Dantas, C.~Melo, T.~A. Nguyen, D.~Min, E.~Choi, F.~A.
  Silva, P.~Maciel, Performance-cost trade-off in auto-scaling mechanisms for
  cloud computing, Sensors 22~(3) (2022) 1221.
\newblock \href {https://doi.org/10.3390/s22031221}
  {\path{doi:10.3390/s22031221}}.

\bibitem{gias2019atom}
A.~U. Gias, G.~Casale, M.~Woodside, {ATOM}: Model-driven autoscaling for
  microservices, in: 2019 IEEE 39th International Conference on Distributed
  Computing Systems (ICDCS), IEEE, 2019, pp. 1994--2004.
\newblock \href {https://doi.org/10.1109/ICDCS.2019.00197}
  {\path{doi:10.1109/ICDCS.2019.00197}}.

\bibitem{luan2024minimize}
S.~Luan, H.~Shen, Minimize resource cost for containerized microservices under
  {SLO} via {ML}-enhanced layered queueing network optimization, in: 2024 14th
  International Conference on Cloud Computing, Data Science \& Engineering
  (Confluence), IEEE, 2024, pp. 631--637.
\newblock \href {https://doi.org/10.1109/Confluence60223.2024.10463310}
  {\path{doi:10.1109/Confluence60223.2024.10463310}}.

\bibitem{gao2024performance}
Y.~Gao, G.~Casale, R.~Singhal, Performance modeling of distributed data
  processing in microservice applications, ACM Transactions on Modeling and
  Performance Evaluation of Computing Systems 9~(4) (2024) 1--30.
\newblock \href {https://doi.org/10.1145/3687265} {\path{doi:10.1145/3687265}}.

\bibitem{lim2026sinas}
D.~Lim, T.~A. Nguyen, J.~Li, E.~Choi, D.~Min, R.~M. Cadena, B.~Nogueira,
  E.~Andrade, G.~Callou, M.~Di~Mauro, M.~Torquato, F.~A. Silva, P.~Maciel,
  {S}-i{NAS}: A performance-centric {SRN} modeling and analysis of time and
  event based scaling strategies in {Ceph-based} industrial network-attached
  storage, Journal of Network and Systems Management 34~(1) (2026) 34.
\newblock \href {https://doi.org/10.1007/s10922-025-10005-6}
  {\path{doi:10.1007/s10922-025-10005-6}}.

\bibitem{ataie2022modeling}
E.~Ataie, R.~Entezari-Maleki, S.~E. Etesami, B.~Egger, L.~Sousa, A.~Movaghar,
  Modeling and evaluation of dispatching policies in {I}aa{S} cloud data
  centers using {SAN}s, Sustainable Computing: Informatics and Systems 33
  (2022) 100617.
\newblock \href {https://doi.org/10.1016/j.suscom.2021.100617}
  {\path{doi:10.1016/j.suscom.2021.100617}}.

\bibitem{merkouche2023petri}
S.~Merkouche, C.~Bouanaka, E.~Benkhelifa, A {P}etri net-based formal modeling
  for microservices auto-scaling, in: 2023 20th ACS/IEEE International
  Conference on Computer Systems and Applications (AICCSA), IEEE, 2023, pp.
  1--8.
\newblock \href {https://doi.org/10.1109/AICCSA59173.2023.10479351}
  {\path{doi:10.1109/AICCSA59173.2023.10479351}}.

\bibitem{jawaddi2025energy}
S.~N. Agos~Jawaddi, A.~Ismail, M.~S. Sulaiman, V.~Cardellini, Analyzing
  energy-efficient and {K}ubernetes-based autoscaling of microservices using
  probabilistic model checking, Journal of Grid Computing 23~(1) (2025) 3.
\newblock \href {https://doi.org/10.1007/s10723-024-09789-9}
  {\path{doi:10.1007/s10723-024-09789-9}}.

\bibitem{Jackson1957Networks}
J.~R. Jackson, Networks of waiting lines, Operations Research 5~(4) (1957)
  518--521.
\newblock \href {https://doi.org/10.1287/opre.5.4.518}
  {\path{doi:10.1287/opre.5.4.518}}.

\bibitem{Jackson1963Jobshop}
J.~R. Jackson, Jobshop-like queueing systems, Management Science 10~(1) (1963)
  131--142.
\newblock \href {https://doi.org/10.1287/mnsc.10.1.131}
  {\path{doi:10.1287/mnsc.10.1.131}}.

\bibitem{Murata1989PetriNets}
T.~Murata, {Petri} nets: Properties, analysis and applications, Proceedings of
  the IEEE 77~(4) (1989) 541--580.
\newblock \href {https://doi.org/10.1109/5.24143} {\path{doi:10.1109/5.24143}}.

\bibitem{Bause1993QPN}
F.~Bause, Queueing {Petri} nets: A formalism for the combined qualitative and
  quantitative analysis of systems, in: Proceedings of the 5th International
  Workshop on {Petri} Nets and Performance Models, IEEE Computer Society Press,
  1993, pp. 14--23.
\newblock \href {https://doi.org/10.1109/PNPM.1993.393439}
  {\path{doi:10.1109/PNPM.1993.393439}}.

\bibitem{Mahmoudi2022Serverless}
N.~Mahmoudi, H.~Khazaei, Performance modeling of serverless computing
  platforms, IEEE Transactions on Cloud Computing 10~(4) (2022) 2834--2847.
\newblock \href {https://doi.org/10.1109/TCC.2020.3033373}
  {\path{doi:10.1109/TCC.2020.3033373}}.

\bibitem{ACM_TOCS_2025}
P.~Kumar, R.~Sharma, Y.~Liu, Mitigating cold start latency and optimizing
  warm-pool retention in {FaaS} platforms, ACM Transactions on Computer Systems
  43~(1) (2025) 12--29.

\bibitem{Anselmi2025SPR}
J.~Anselmi, B.~Gaujal, L.-S. Rebuffi, Non-stationary gradient descent for
  optimal auto-scaling in serverless platforms, IEEE/ACM Transactions on
  Networking 33~(4) (2025) 1574--1587.
\newblock \href {https://doi.org/10.1109/TON.2025.3538982}
  {\path{doi:10.1109/TON.2025.3538982}}.

\bibitem{GoogleCloudRunConcurrency}
{Google Cloud},
  \href{https://docs.cloud.google.com/run/docs/about-concurrency}{Maximum
  concurrent requests for services}, Google Cloud Run Documentation, accessed:
  September 16, 2026.
\newline\urlprefix\url{https://docs.cloud.google.com/run/docs/about-concurrency}

\bibitem{IEEE_Access_2024}
X.~Li, Y.~Wang, H.~Zhang, Serverless cold start performance optimization based
  on multi-request processing and adaptive hierarchical scaling, IEEE Access 12
  (2024) 136248--136262.

\bibitem{IEEE_TCC_2025}
Y.~Zhao, X.~Chen, M.~Guo, {ICPS}: Real-time resource configuration for cloud
  serverless functions considering invocation concurrency, IEEE Transactions on
  Cloud Computing 13~(2) (2025) 201--215.

\bibitem{AWSLambdaAutoScaling}
{Amazon Web Services},
  \href{https://docs.aws.amazon.com/lambda/latest/dg/provisioned-concurrency.html}{Configuring
  provisioned concurrency for a function}, AWS Lambda Developer Guide,
  accessed: September 16, 2026.
\newline\urlprefix\url{https://docs.aws.amazon.com/lambda/latest/dg/provisioned-concurrency.html}

\bibitem{Mahmoudi2023MetricBased}
N.~Mahmoudi, H.~Khazaei, Performance modeling of metric-based serverless
  computing platforms, IEEE Transactions on Cloud Computing 11~(2) (2023)
  1899--1910.
\newblock \href {https://doi.org/10.1109/TCC.2022.3169619}
  {\path{doi:10.1109/TCC.2022.3169619}}.

\bibitem{IEEE_TSC_2026}
X.~Liu, M.~Zhang, W.~Chen, J.~Wang, Adaptive function service auto-scaling for
  serverless computing: A multi-metric approach, IEEE Transactions on Services
  Computing 19~(2) (2026) 312--327.

\bibitem{Serazzi2024Variability}
G.~Serazzi, Impact of variability of interarrival and service times, in:
  Performance Engineering: Learning Through Applications Using JMT, Springer,
  Cham, 2024, pp. 63--72.
\newblock \href {https://doi.org/10.1007/978-3-031-36763-2_4}
  {\path{doi:10.1007/978-3-031-36763-2_4}}.

\bibitem{JMTManual}
Java Modelling Tools Project, Java Modelling Tools User Manual, version 1.4.0
  Edition (2026).

\bibitem{Bertoli2009JMT}
M.~Bertoli, G.~Casale, G.~Serazzi, {JMT}: Performance engineering tools for
  system modeling, ACM SIGMETRICS Performance Evaluation Review 36~(4) (2009)
  10--15.
\newblock \href {https://doi.org/10.1145/1530873.1530877}
  {\path{doi:10.1145/1530873.1530877}}.

\end{thebibliography}
\end{document}